\documentclass{jfm}
\usepackage{amsmath}
\usepackage{amsfonts}
\usepackage{graphicx}
\usepackage{hyperref}
\usepackage{subcaption}
\usepackage{cleveref}
\usepackage{mathrsfs}
\usepackage{tikz}
\usepackage{mathtools}
\usetikzlibrary{positioning}
\newcommand{\tab}{\hspace*{3em}}
\graphicspath{{./figures/}}
\newcommand{\vol}{\mathop{\ooalign{\hfil$V$\hfil\cr\kern0.08em--\hfil\cr}}\nolimits}
\shorttitle{ECS for grooved PCF}
\shortauthor{S. B. Vadarevu, A. Sharma and B. Ganapathisubramani}

\title{Exact coherent states for grooved Couette flows}

\author{Sabarish B. Vadarevu
  \corresp{\email{sabarish.vadarevu@unimelb.edu.au}},
  A. Sharma
 \and B. Ganapathisubrmani}

\affiliation{Aerodynamics and Flight Mechanics group, University of Southampton,
Southampton, Hampshire SO17 1BJ, UK}
\begin{document}

\maketitle

\begin{abstract}
    Recent progress indicates that highly symmetric recurring solutions of the Navier-Stokes equations such as equilibria and periodic orbits provide a skeleton for turbulence dynamics in state-space. Many of these solutions have been found for flat-walled plane Couette, channel, and pipe flows. Rough-walled flows are of great practical significance, yet no recurring solutions are known for these flows. 
    We present a numerical homotopy method to continue solutions from flat-walled plane Couette flow (PCF) to grooved PCF, demonstrated here at a Reynolds number of 400, to act as a starting point for similar continuation to rough-walled flows. Loss of spanwise homogeneity in grooved PCF reduces continuous families of solutions (identical up to translational shifts) in flat-walled Couette flow to multiple, discrete families in grooved Couette flow; this can manifest in turbulence as spatially anchored exact coherent structures near the wall, so that turbulent statistics reflect symmetry-restricted structure of exact recurring solutions. 
    Furthermore, the vortex-streak structures characteristic of these equilibria are squeezed out of the grooves when the groove-wavelength is smaller than the characteristic spanwise size of the structures. This produces reduced shear stress at the wall even at the low Reynolds numbers considered, and the mechanism is consistent with the drag reduction observed in some riblet-mounted turbulent flows. 
\end{abstract}


\section{Introduction}
For most of the twentieth century, turbulence was studied through the lens of statistics despite knowledge of the existence of recurring structures. Townsend's attached eddy hypothesis \citep{townsend1980structure} was an early attempt to describe turbulence in terms of coherent structures and relate their scaling to observed statistics. These eddies were suspected to be the building blocks of turbulence \citep{perry1995wall}, but they were ad-hoc structures without basis in the Navier-Stokes equations (NSE). Advances in computational power allowed computation of so-called exact coherent structures (ECS), starting with plane Couette flow (PCF) equilibria of \cite{nagata1990three}. Minimal channel DNS of \cite{jimenez1991minimal} confirmed the significance of the interaction between streamwise streaks and vortices seen in the equilibria. \citet{waleffe1997self} provided a detailed picture of the recurring motions that occur in wall-bounded turbulence; the linear instability and non-linear breakdown of streaks produces vortices, which in turn feed into the streaks. 

In the past decade, many exact solutions of the Navier Stokes equations have been found and their connections in state-space have been mapped \citep[for instance ][]{gibson2008visualizing,eckhardt2008dynamical,kawahara2012significance,park2015exact,willis2015relative}. A plausible description of the underlying dynamics at asymptotically high Reynolds numbers has also been proposed \citep{hall1991strongly,hall2010streamwise} highlighting the inviscid mechanism of vortex-wave interactions. Today, these solutions --- equilibria, travelling waves, periodic orbits, and relative periodic orbits --- are understood to shepherd turbulent trajectories in state-space. 

While rapid progress has been made towards understanding turbulence dynamics in terms of exact recurrent solutions of the NSE, such efforts are yet to involve wall-roughness, despite the practical significance of rough-walled turbulence. Studying the variation in the structure and statistics of exact solutions with changing wall geometry (from smooth to rough) can offer insight into the interactions of wall-roughness with turbulent flow. Such an approach can complement the work to date on rough-walled turbulence \citep{jimenez2004turbulent,flack2014roughness,squire2016comparison}, where turbulence structure and statistics are examined using experimental and numerical methods, to provide a quantitative basis for characterizing surface topologies and relating them to measures such as equivalent sand roughness. 

We consider the special case of riblets (longitudinal grooves) to serve as a starting point for the extension of ECS from smooth-walled to rough-walled geometries; we continue the equilibria of \cite{nagata1990three}, which were computed independently and hosted on the website www.channelflow.org by \cite{channelflow}. In addition to the geometrical simplicity, riblet-mounted flows are particularly interesting because of the drag reduction observed for certain arrangements \citep{walsh1990effect,dean2010shark,bannier2016riblets}. This drag reduction is accompanied by a localization of the vortex-streak structure near the wall \citep{lee2001flow}. We expose the drag reduction and localization to be related to the influence of riblets on symmetric ECS. 

The original solutions of \citet{nagata1990three} were obtained using homotopy from Taylor-Vortex flow to PCF. We use homotopy to extend the equilibria of \citet{nagata1990three} from flat-walled PCF (smooth-walled PCF) to riblet-mounted PCF using a domain transformation method. Henceforth, we refer to flat-walled PCF as flat PCF, and riblet-mounted PCF as grooved PCF. We also distinguish the base flow $U=y$ (and solutions obtained from continuing this to grooved PCF) as the laminar solution, while the 3-dimensional solutions of \cite{nagata1990three} and their continuations are referred to as equilibria. 


\subsection{Flow symmetries}\label{sec:symmetries-smooth}
\subsubsection{Continuous symmetries}\label{sec:symmetries-continuous}
Smooth-walled PCF involves infinite, parallel plates moving in opposite directions. This allows for solutions that are invariant under certain symmetry transformations; we follow the notation of \citet{gibson2009equilibrium} to represent these symmetries. We use non-dimensional coordinates $(x,y,z)$ along the streamwise, wall-normal, and spanwise directions respectively, non-dimensionalized by half of the distance between the walls. 

Homogeneity in the streamwise and spanwise directions leads to admission of solutions that are invariant under continuous translation in the $x-z$ plane; the symmetry transformations for these translations become the continuous two-parameter group $SO(2)_x \times SO(2)_z$, denoted by  
\begin{equation}
	\tau(l_x,l_z) \lbrack u,v,w \rbrack (x,y,z) = \lbrack u,v,w \rbrack (x+l_x, y, z+l_z).
\end{equation}

This homogeneity also means that spatially periodic solutions can be sought, so that computations can be carried out over small domain sizes using the Fourier spectral method for discretization. Different box sizes produce different solutions; however, provided that the box-sizes are within certain ranges --- of the order of the channel half-height --- these solutions all seem to contain the vortex-streak structure. 

While the homogeneity allows for relatively inexpensive, spatially periodic solutions, it also produces vast families of solutions that are dynamically similar. That is, every equilibrium $\lbrack u^*,v^*,w^* \rbrack$ is part of a continuous family of solutions spanned by $\tau(l_x,l_z) \lbrack u^*,v^*,w^* \rbrack$ for arbitrary $l_x,l_z$, and can also have dynamically similar travelling waves associated with the family. Presence of such families could obscure our view of turbulent flows. \citet{willis2013revealing} developed a projection method called the ``method of slices'' to identify families of dynamically similar solutions with one representative solution for each family, thus reducing the number of solutions that would have to be considered, say, when developing low-order models for control. Specifically, flat-walled plane shear flows admit travelling wave solutions and relative periodic orbits, and the method of slices relates such solutions to dynamically similar equilibria and periodic orbits by projecting the former onto a hyperplane using a pre-defined template. Grooved flows, however, are inhomogeneous in the direction transverse to the grooves. The consequences of this loss of homogeneity are explored in \cref{sec:localization}. 

\subsubsection{Discretization}\label{sec:discretization-smooth}

Owing to the existence of spatially periodic solutions for channel flows and PCF, the flowfield can be discretized using the Fourier spectral method in the wall-parallel directions. A periodic field $\phi(x,y,z,t)$ for a periodic box of streamwise and spanwise lengths $L_x$ and $L_z$ is written as a sum of Fourier modes,
\begin{equation}
    \phi(x,y,z,t) = \sum\limits_{k \in \mathbb{Z}} \sum\limits_{l \in \mathbb{Z}} \sum\limits_{m \in \mathbb{Z}} \hat{\phi}_{k,l,m}(y,t) e^{i(l\alpha x + m \beta z )},
\end{equation}
where $\alpha,\beta = 2\pi/L_x, 2\pi/L_z$. The Fourier modes are truncated to some $|l| \leq L$ and $|m| \leq M$ to obtain a finite-dimensional representation of the field variable in $x$ and $z$. Henceforth, hats on the coefficients of Fourier modes are dropped for convenience, with the understanding that integer-subscripted symbols represent Fourier coefficients. 

The Fourier coefficients are functions of the wall-normal coordinate and time. Chebyshev collocation method is used for discretization along the wall-normal direction. The time dependence is usually accounted for by marching forward from an initial condition; but in the steady-state solver used in this work, this dependence is dropped.

\subsubsection{Discrete symmetries}\label{sec:discrete-symmetries}
Smooth PCF also admits discrete symmetries, which are restricted by the counter-moving walls to reflection across $x-y$ planes ($\sigma_z$), rotation about the $z$-axis ($\sigma_x$), and point-wise inversion about the origin ($\sigma_{xz}$), which is the product of the reflection and rotation. These symmetry transformations generate a discrete dihedral group $D_1 \times D_1 = \{ e, \sigma_x, \sigma_z, \sigma_{xz} \}$, where
\begin{equation}
	\begin{aligned}
		\sigma_x \lbrack u,v,w \rbrack (x,y,z) &= \lbrack -u,-v,w \rbrack (-x, -y, z),\\
		\sigma_z \lbrack u,v,w \rbrack (x,y,z) &= \lbrack u,v,-w \rbrack (x, y, -z),\\
		\sigma_{xz} \lbrack u,v,w \rbrack (x,y,z) &= \lbrack -u,-v,-w \rbrack (-x, -y, -z),\\
	\end{aligned}
\end{equation}
and $e$ is the identity transformation. 
The equations of flat PCF are invariant under the group $\Gamma = D_{1,x} \ltimes SO(2)_x \times D_{1,z} \ltimes SO(2)_z$, where $\ltimes$ stands for a semi-direct product, $x$ subscripts indicate translations along $x$ and sign changes in $x$ and $y$, and $z$ subscripts indicate translations along $z$ and sign changes in $z$. 

The laminar solution (base flow, $U=y$) for the flat PCF system is invariant under every symmetry transformation in $\Gamma$. Equilibria, such as those of \citet{nagata1990three}, are invariant under fewer symmetries. Streamwise travelling waves cannot exist when $\sigma_x$ is imposed, spanwise travelling waves cannot exist when $\sigma_z$ is imposed, and neither can exist when $\sigma_{xz}$ is imposed. More complex solutions appear as these symmetries are relaxed, with the fully turbulent state unlikely to satisfy any of the symmetries in $\Gamma$ (although discrete translational symmetries are imposed in simulations in the form of periodic boxes). The reader is referred to \citet{gibson2009equilibrium,halcrow2008charting} for a comprehensive treatment of invariance under discrete symmetry groups. They identify one particular order-4 group, $S=\{e, s_1, s_2, s_3\}$, to be important to finding equilibria of flat PCF, where
\begin{equation}\label{symmetries-smooth}
	\begin{aligned}
		s_1 &= \tau(L_x/2, 0) \sigma_z,\\
		s_2 &= \tau(L_x/2, L_z/2) \sigma_x, \\
		s_3 &= \tau(0, L_z/2) \sigma_{xz},
	\end{aligned}
\end{equation}
are products of half-cell shifts and the discrete symmetries of $D_{1,x} \times D_{1,z}$. The solutions used for continuation in the present work are invariant under the symmetries of $S$.

Due to the high-dimensionality of the state-space of flat PCF, and fluid flows in general, a large number of exact invariant solutions exist; in fact, an infinite number of solutions exist due to the continuous translational invariance of the governing equations. A small set of dynamically important solutions has to be identified in order to realistically describe and predict turbulence. The symmetry groups $S$, $S \times \{ e, \sigma_{xz}\}$, and other specific subgroups of $\Gamma$ offer one way of reducing the infinite stationary points of the NSE to a small set, while the method of slices described earlier offers a complementary way to constraining continuous symmetries. 

\section{Methodology}

\subsection{Domain transformation for grooved PCF}
A general case of wall roughness periodic in $(L_x, L_z) = (2 \pi/\alpha, 2\pi/\beta)$ can be expressed as a Fourier series,
\begin{equation}\label{physical-domain-general}
		y_{top} = 1 + \sum\limits_{l,m} A^t_{l,m} e^{i(l\alpha x + m\beta z + \phi^t_{l,m})}, \quad
		y_{bottom} = -1 + \sum\limits_{l,m} A^b_{l,m} e^{i(l\alpha x + m\beta z + \phi^b_{l,m})}.
\end{equation}

This grooved geometry can be mapped to a flat-walled geometry using a domain transformation, $Y = -1 + 2(y-y_{bottom})/(y_{top}- y_{bottom})$ (for similar usage, see \cite{kasliwal2012modelling,moradi2013maximization}), so that the Fourier spectral method is easily applied in the transformed domain. Mapping partial derivatives from the transformed domain to the physical domain of \cref{physical-domain-general} is, in general, rather cumbersome. However, when $A^t_{l,m} = A^b_{l,m}$ and $\phi^t_{l,m} = \phi^b_{l,m}$, i.e., when the width between the top and bottom walls remains constant, the required domain transformation is
\begin{equation}
    Y =  y - \sum\limits_{l,m} A_{l,m} e^{i(l\alpha x + m\beta z + \phi_{l,m})},
\end{equation}
which produces simpler relations between the partial derivatives in the physical domain of \cref{physical-domain-general} and the transformed domain.

We investigate the special case of longitudinal grooves ($\alpha = 0$), with the additional simplification that $\phi_{m} = 0$. The top and bottom walls of the PCF are given by 
\begin{equation}\label{physical-domain}
    y_{walls} = \pm 1 + \sum\limits_m A_m \cos(m\beta z),
\end{equation}
and the grooved PCF is mapped to a flat PCF geometry as
\begin{equation}\label{domain-transformation}
		X = x, \tab
		Y = T(y,z) = y - \sum\limits_m A_m \cos(m\beta z),\tab
		Z  = z.
\end{equation}

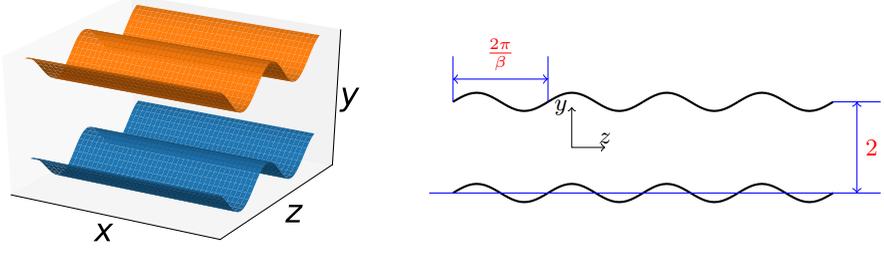
\begin{figure}
	\centering
    \begin{subfigure}{0.47\textwidth}
		\includegraphics[width=0.95\textwidth]{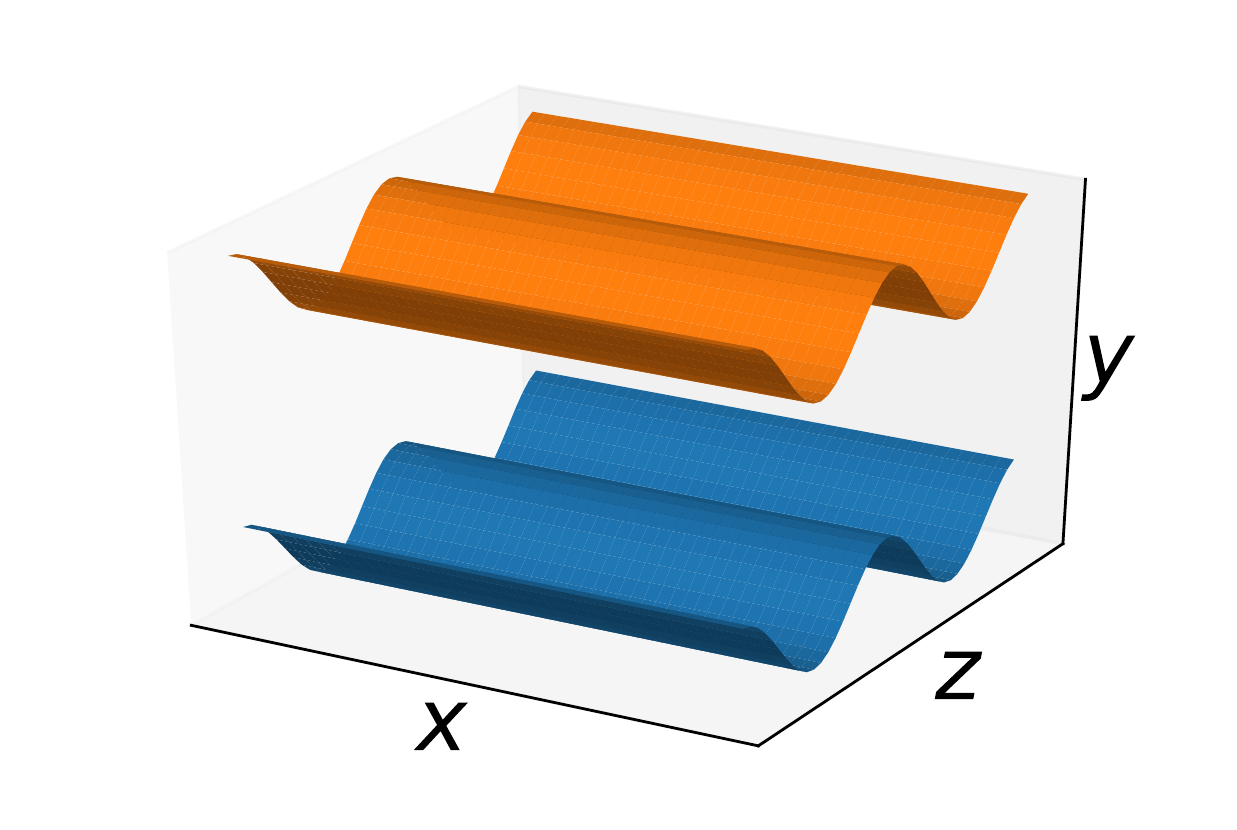}
    \end{subfigure}
    \begin{subfigure}{0.47\textwidth}
        \begin{tikzpicture}[xscale=0.2,yscale=0.6]
            \draw [thick, domain=0:2*pi] plot (\x, {0.2*sin(\x r)});\draw [thick, domain=2*pi:4*pi] plot (\x, {0.2*sin(\x r)}); \draw [thick, domain=4*pi:6*pi] plot (\x, {0.2*sin(\x r)});\draw [thick, domain=6*pi:8*pi] plot (\x, {0.2*sin(\x r)});
            \draw [thick, domain=0:2*pi] plot (\x, {2+0.2*sin(\x r)});\draw [thick, domain=2*pi:4*pi] plot (\x, {2+0.2*sin(\x r)}); \draw [thick, domain=4*pi:6*pi] plot (\x, {2+0.2*sin(\x r)});\draw [thick, domain=6*pi:8*pi] plot (\x, {2+0.2*sin(\x r)});
        \draw [<->,very thin] (3.2*pi,1) -- (2.5*pi,1) -- (2.5*pi,0.7*pi/2.5+1); 
        \node [above] at (3.2*pi,1-0.1) {$z$};\node [left] at (2.5*pi+0.3,0.7*pi/2.5+1) {$y$};
        \draw [blue, thin](-pi/2,0) -- (9*pi,0);\draw [blue, thin](8*pi,2) -- (9*pi,2);
        \draw [<->,blue,thin] (17*pi/2,0) -- (17*pi/2,2);
        \node [right,red] at (17*pi/2,1) {2};
        \draw [blue, thin](0,2) -- (0,3);\draw [blue, thin](2*pi,2) -- (2*pi,3);
        \draw [<->,blue,thin] (0,2.5) -- (2*pi,2.5);
        \node [above,red] at (pi,2.5) {$\frac{2\pi}{\beta}$};
        \end{tikzpicture}
    \end{subfigure}
        \caption[Grooved PCF with longitudinal grooves]{Couette flow between infinite plates with longitudinal grooves (with $A_{m\neq 1} = 0$). The grooves are arranged ``in-phase'' so that the distance between the plates remains constant. \label{fig:groovedPCF}}
\end{figure}

The partial derivatives in the two domains relate as
\begin{equation}\label{partial-derivatives}
		\partial_x = \partial_X,\tab
		\partial_y = \partial_Y,\tab
        \partial_z = \partial_Z + \bigg( \sum\limits_m m \beta A_m \sin(m\beta z)\bigg) \partial_Y.
\end{equation}

\subsection{Iterative solver}
A primitive variable formulation is used in a steady-state solver to extend the equilibria of \citet{nagata1990three} to grooved PCF; the solver is outlined in appendix \ref{app:solver}. The velocities $u,v,w$ along $x,y,z$ are non-dimensionalized by the speed of the walls; the reference frame is chosen such that the walls have velocities $\pm U_{walls}$ at $y = \pm 1$. The pressure is non-dimensionalized by $\rho U^2_{walls}$, where $\rho$ is the density of the fluid. The equations to be solved are
\begin{equation}\label{governing-equations}
	\begin{aligned}
		u \partial_x u + v\partial_y u + w \partial_z u + \partial_x p - \frac{1}{Re}\Delta u &= 0,\\
		u \partial_x v + v\partial_y v + w \partial_z v + \partial_y p - \frac{1}{Re}\Delta v &= 0,\\
		u \partial_x w + v\partial_y w + w \partial_z w + \partial_z p - \frac{1}{Re}\Delta w &= 0,\\
		\partial_x u + \partial_y v + \partial_z w &=0,
	\end{aligned}
\end{equation}
where $\Delta = \partial_{xx} + \partial_{yy} + \partial_{zz}$. 
The boundary conditions are 
\begin{equation}\label{boundary-conditions}
	\begin{gathered}
		u(y=y_{top}) = 1, \tab u(y=y_{bottom}) = -1, \\
		v(y=y_{walls}) = 0, \tab w(y=y_{walls}) = 0,\tab
		\{\nabla \cdot \mathbf{u}\}(y=y_{walls}) = 0.
	\end{gathered}
\end{equation}
The Reynolds number $Re$ is based on the dimensional half-width and the dimensional speed of the walls. 

The velocities and pressure are mapped from the grooved PCF domain with walls given in \cref{physical-domain} to a flat PCF domain according to \cref{domain-transformation}; the boundary conditions in the transformed domain are  
\begin{equation}\label{boundary-conditions-transformed}
    \begin{aligned}
        u(Y=\pm 1) &= \pm 1, \tab v(Y=\pm 1) &= 0,\\
        w(Y=\pm 1) &= 0, \tab \{\nabla \cdot \mathbf{u} \} (Y=\pm 1) &= 0.
    \end{aligned}
\end{equation}

The fields are discretized using the Fourier spectral method along the streamwise and spanwise directions and the Chebyshev collocation method along the wall-normal direction in the transformed domain $(X,Y,Z)$. \Cref{governing-equations} is discretized by writing the partial derivatives in $x,y,z$ in terms of partial derivatives in $X,Y,Z$ using \cref{partial-derivatives}. A major consequence of the domain transformation is that the Fourier coefficients of the spanwise derivatives involve multiple Fourier modes:
\begin{equation}\label{partial-derivatives-modes}
    \begin{gathered}
		\{ \partial_x u\}_{l,m} = il \alpha u_{l,m},\quad
		\{ \partial_y u\}_{l,m} = \partial_Y (u_{l,m}),\\
		\{ \partial_z u\}_{l,m} = im \beta u_{l,m} + \frac{-i\beta}{2} \sum\limits_q q A_q u_{l,m-q},
    \end{gathered}
\end{equation}
where $u_{l,m}$ is the $Y$-dependent Fourier coefficient for mode $(l\alpha, m\beta)$. This additional inter-modal interaction contains all the information due to the grooved walls, and is the trade-off for the simplified boundary conditions of \cref{boundary-conditions-transformed}.

The discretized equations are solved using the Newton-Raphson method (see appendix \ref{app:solver}). A full-rank matrix inversion (SVD-based) is used in the iterative correction stpdf. The iterations are initialized with exact solutions for flat PCF: EQ1 (lower branch) and EQ2 (upper branch) of \citet{nagata1990three}, taken from the solutions database of \citet{channelflow}. Our steady-state solver defines instantaneous time-derivative and divergence of the velocity field as the residual ($\mathbf{r}$), and the residual norm is defined as the integral
\begin{equation}\label{residual}
    ||\mathbf{r}|| = \bigg\lbrack \int_{x=0}^{2\pi/\alpha} \int_{z=0}^{2\pi/\beta} \int_{y=y_{bottom}}^{y_{top}} \big\{\dot{u}^2 + \dot{v}^2 +\dot{w}^2 + (\nabla \cdot \boldsymbol{u})^2 \big\} \mathrm{d}y \mathrm{d}z \mathrm{d}x \bigg\rbrack^{1/2}.
\end{equation}

Solutions for grooved PCF are computed with 35 wall-normal Chebyshev nodes and 825 Fourier modes: $|l| \leq 14$ and $ |m|\leq 16$ along the streamwise and spanwise directions respectively. Spatial accuracy of the solutions is defined as the residual norm of the solutions when extrapolated on a grid with 70 Chebyshev nodes and 3185 Fourier modes: $|l| \leq 28$ and $|m| \leq 32$.

\section{Results and Discussion}
\begin{figure}
	\begin{subfigure}{0.47\textwidth}
		\begin{tikzpicture}
			\node (img) {\includegraphics[width=\textwidth]{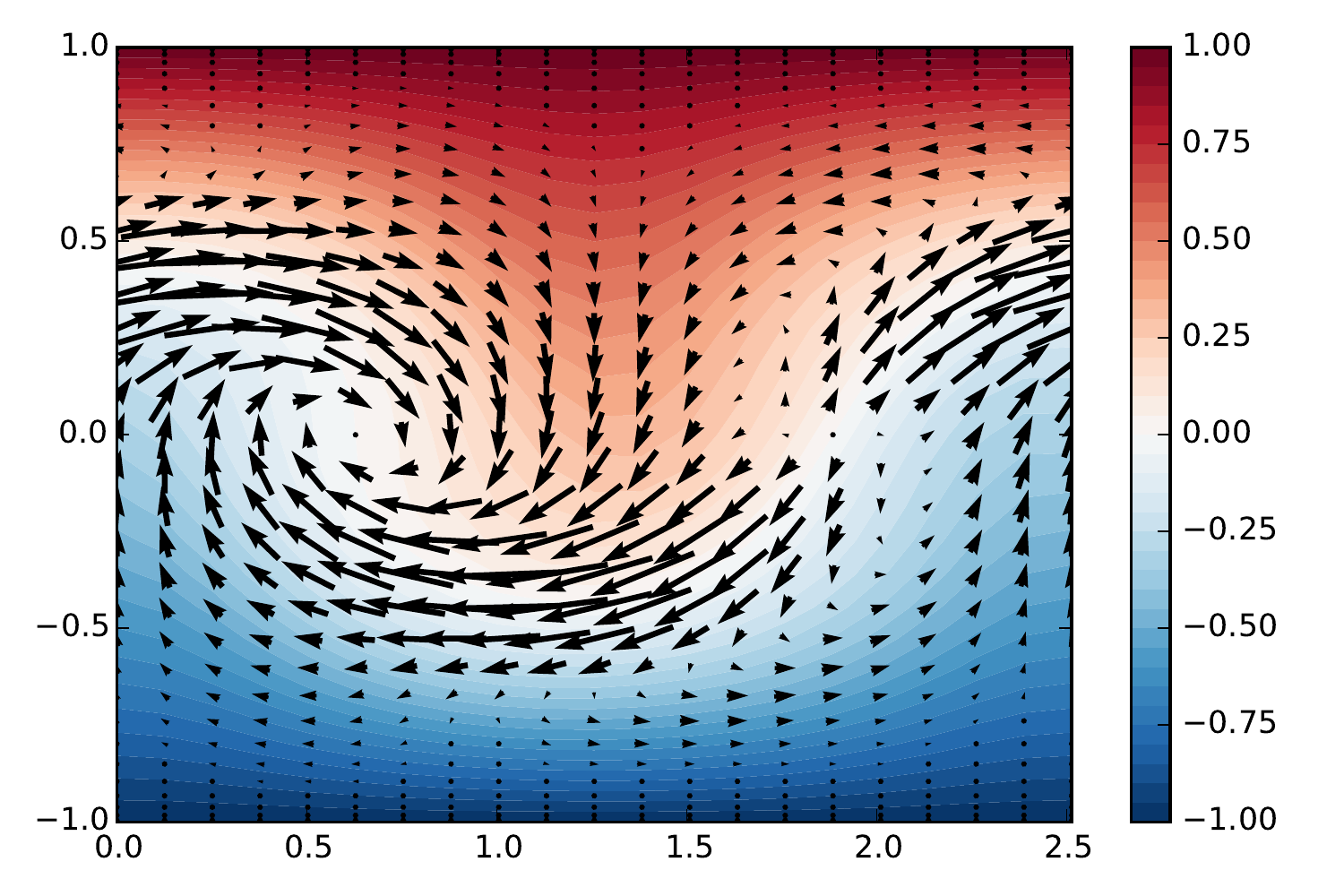}};
			\node[below=of img, node distance=0cm, xshift=-0.2cm,yshift=1.2cm]{$z$};
			\node[left=of img, node distance=0cm, rotate=90, anchor=center, yshift=-1.1cm]{$y$};
		\end{tikzpicture}
		\caption{EQ1; flat PCF\label{fig:lower-flat}}
	\end{subfigure}
	\begin{subfigure}{0.47\textwidth}
		\begin{tikzpicture}
			\node (img) {\includegraphics[width=\textwidth]{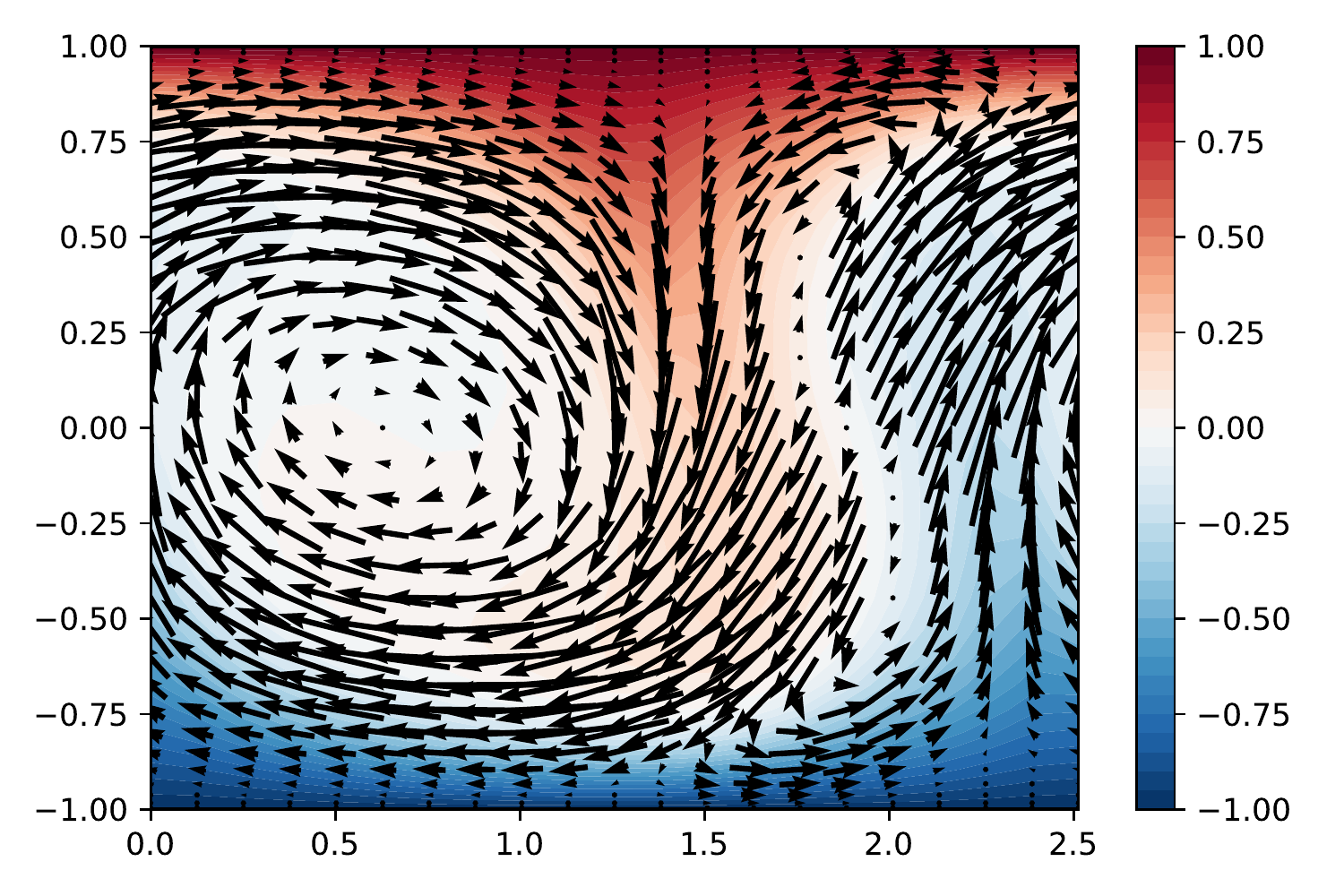}};
			\node[below=of img, node distance=0cm, xshift=-0.2cm,yshift=1.2cm]{$z$};
			\node[left=of img, node distance=0cm, rotate=90, anchor=center, yshift=-1.1cm]{$y$};
		\end{tikzpicture}
		\caption{EQ2; flat PCF\label{fig:upper-flat}}
	\end{subfigure}\\
	\begin{subfigure}{0.47\textwidth}
		\begin{tikzpicture}
			\node (img) {\includegraphics[width=\textwidth]{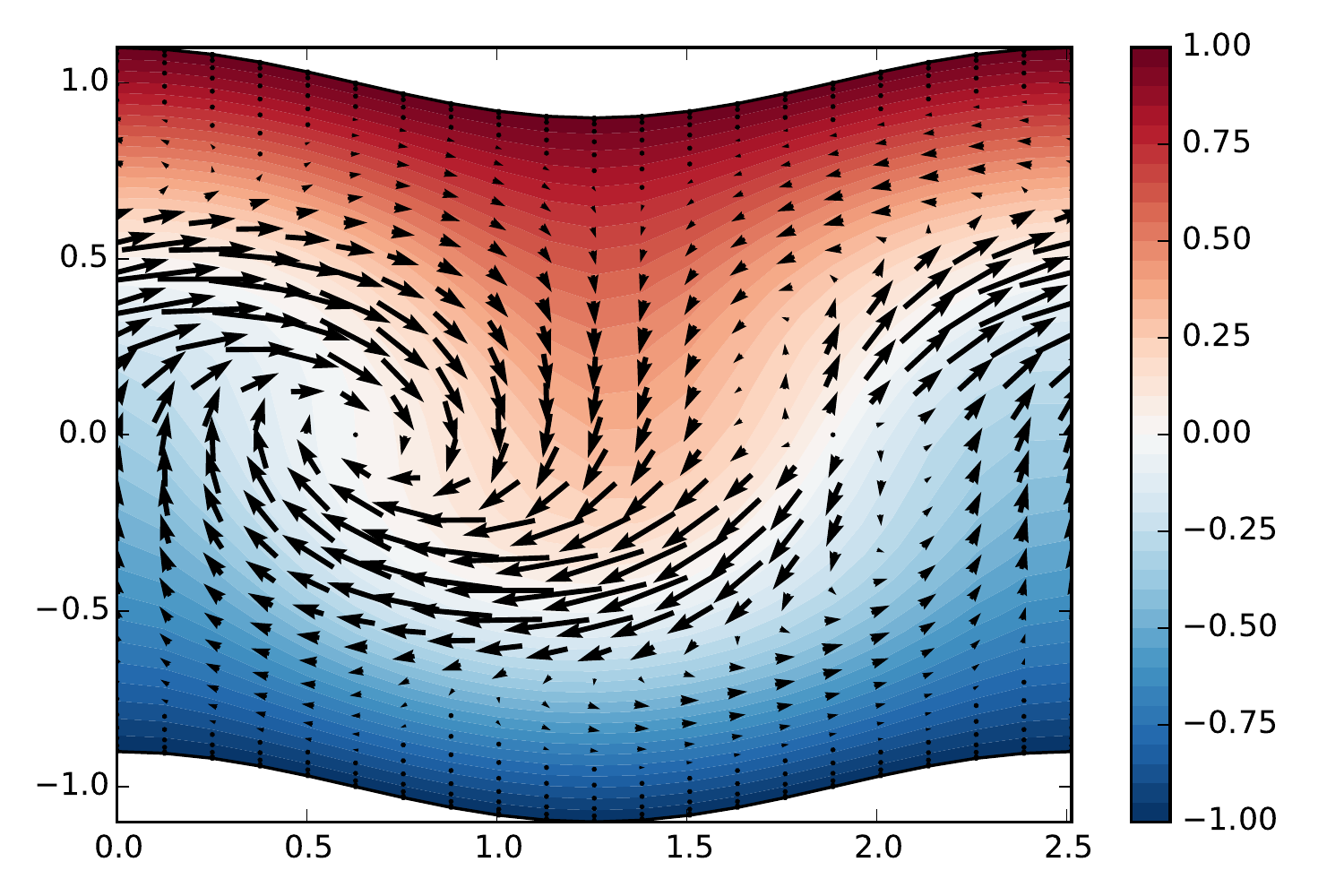}};
			\node[below=of img, node distance=0cm, xshift=-0.2cm,yshift=1.2cm]{$z$};
			\node[left=of img, node distance=0cm, rotate=90, anchor=center, yshift=-1.1cm]{$y$};
		\end{tikzpicture}
		\caption{EQ1; grooved PCF with $A_1 = 10\%$ \label{fig:lower-grooved}}
	\end{subfigure}
	\begin{subfigure}{0.47\textwidth}
		\begin{tikzpicture}
			\node (img) {\includegraphics[width=\textwidth]{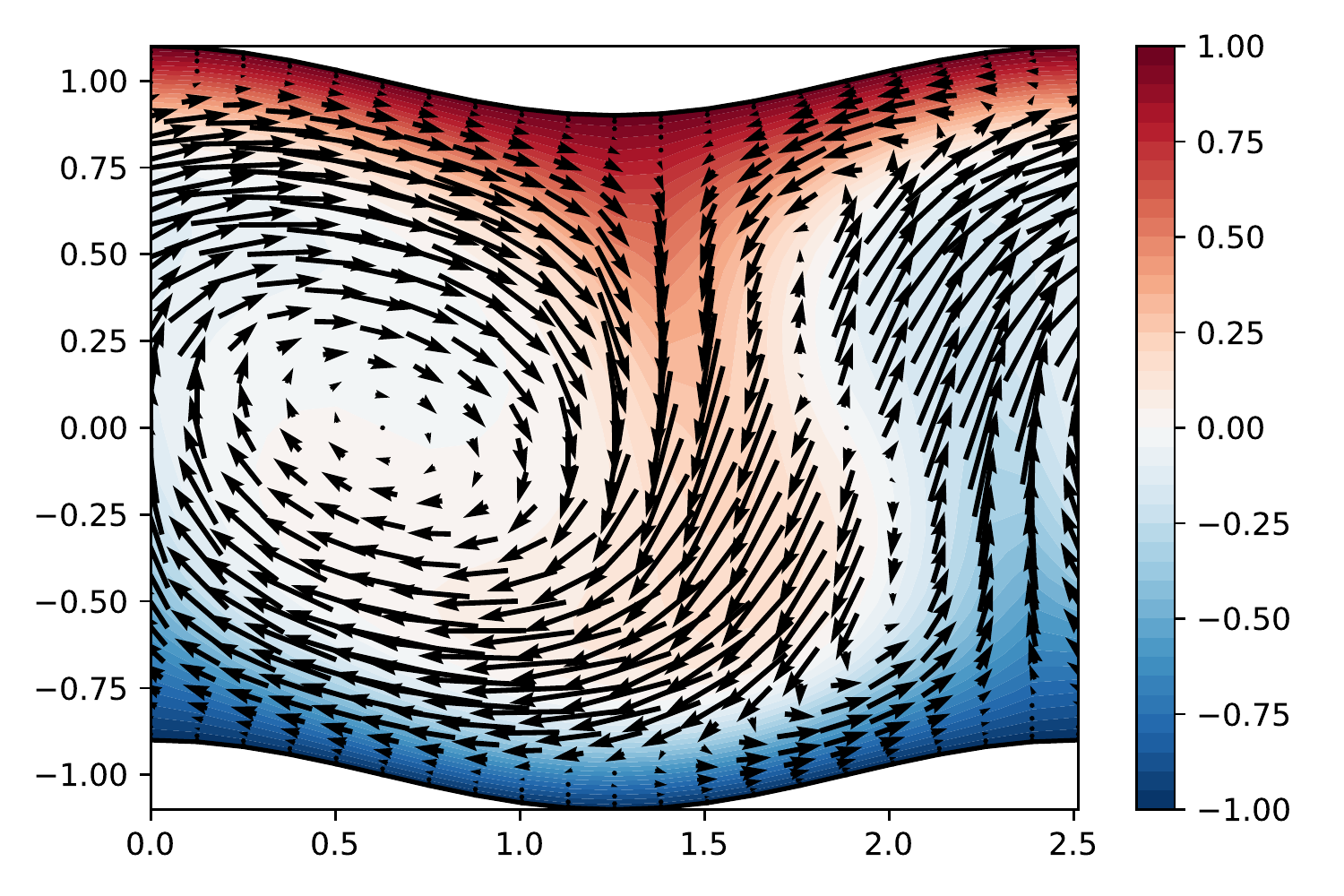}};
			\node[below=of img, node distance=0cm, xshift=-0.2cm,yshift=1.2cm]{$z$};
			\node[left=of img, node distance=0cm, rotate=90, anchor=center, yshift=-1.1cm]{$y$};
		\end{tikzpicture}
		\caption{EQ2; grooved PCF with $A_1 = 10\%$ \label{fig:upper-grooved}}
	\end{subfigure}
	\caption{Velocity on a cross-stream plane ($x=0$) for flat PCF (a-b) and grooved PCF (c-d). Contours show streamwise velocity and quiver arrows show cross-stream velocities.\label{fig:velocity}}
\end{figure}
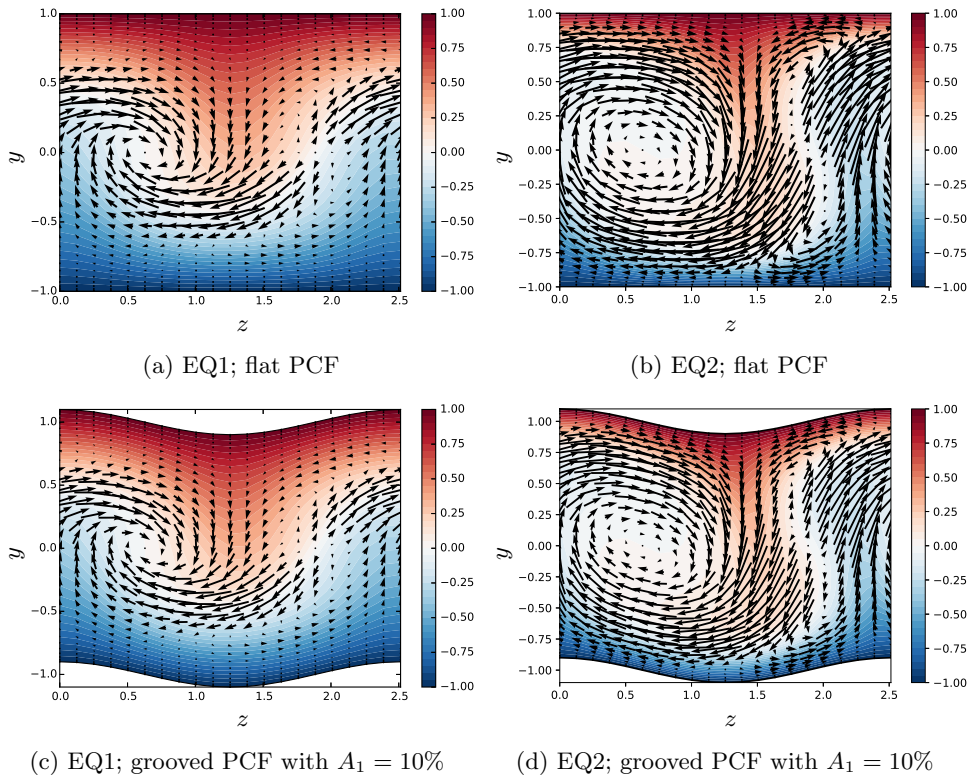

\begin{figure}
    \begin{subfigure}{0.98\textwidth}
	\centering
    \includegraphics[width=\textwidth]{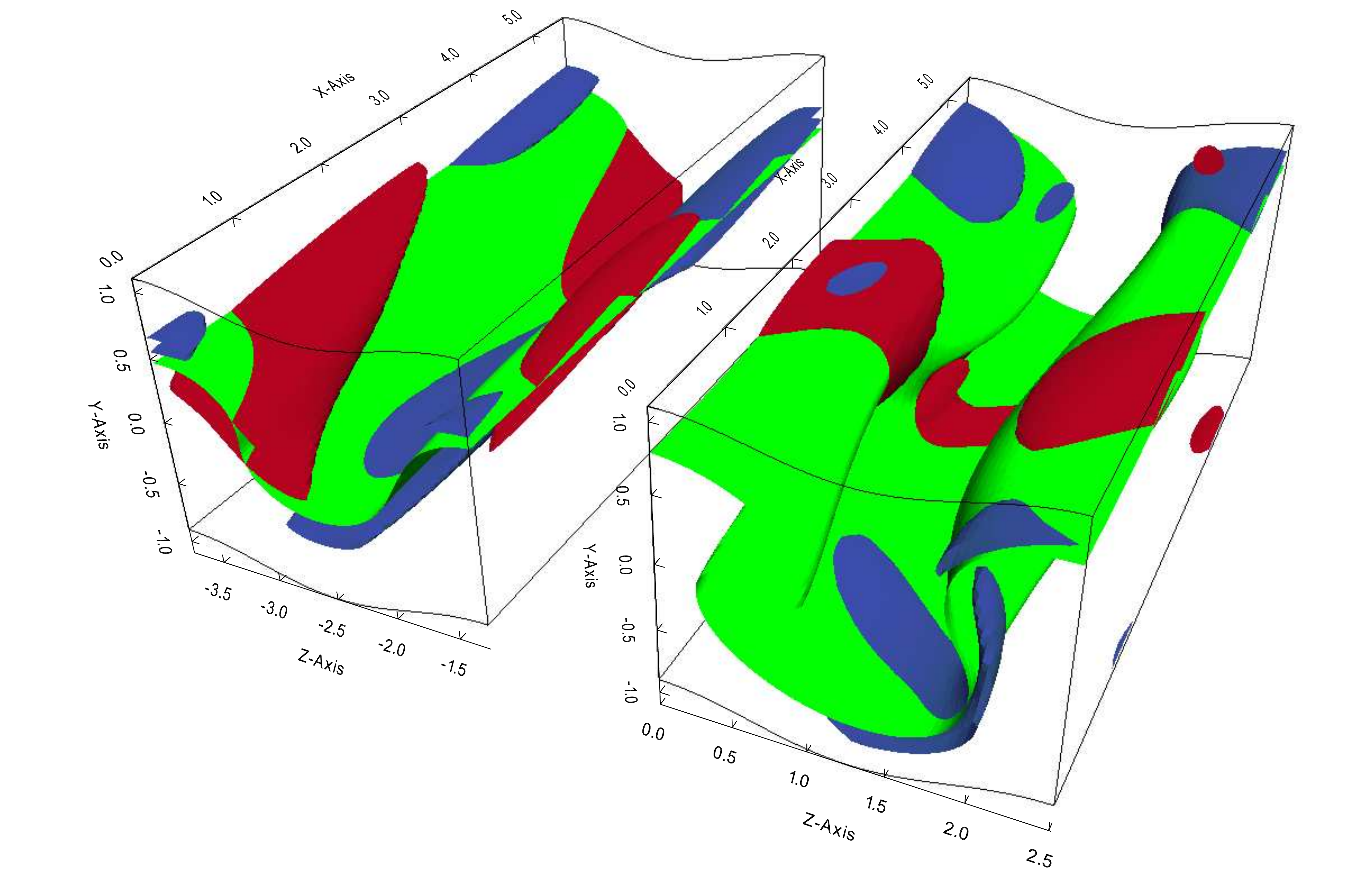}
    \caption{EQ1 (left) and EQ2 (right) at $A = 0.1$}
    \end{subfigure}\\
	\begin{subfigure}{0.44\textwidth}
	\centering
		\begin{tikzpicture}
			\node (img) {\includegraphics[width=0.9\textwidth]{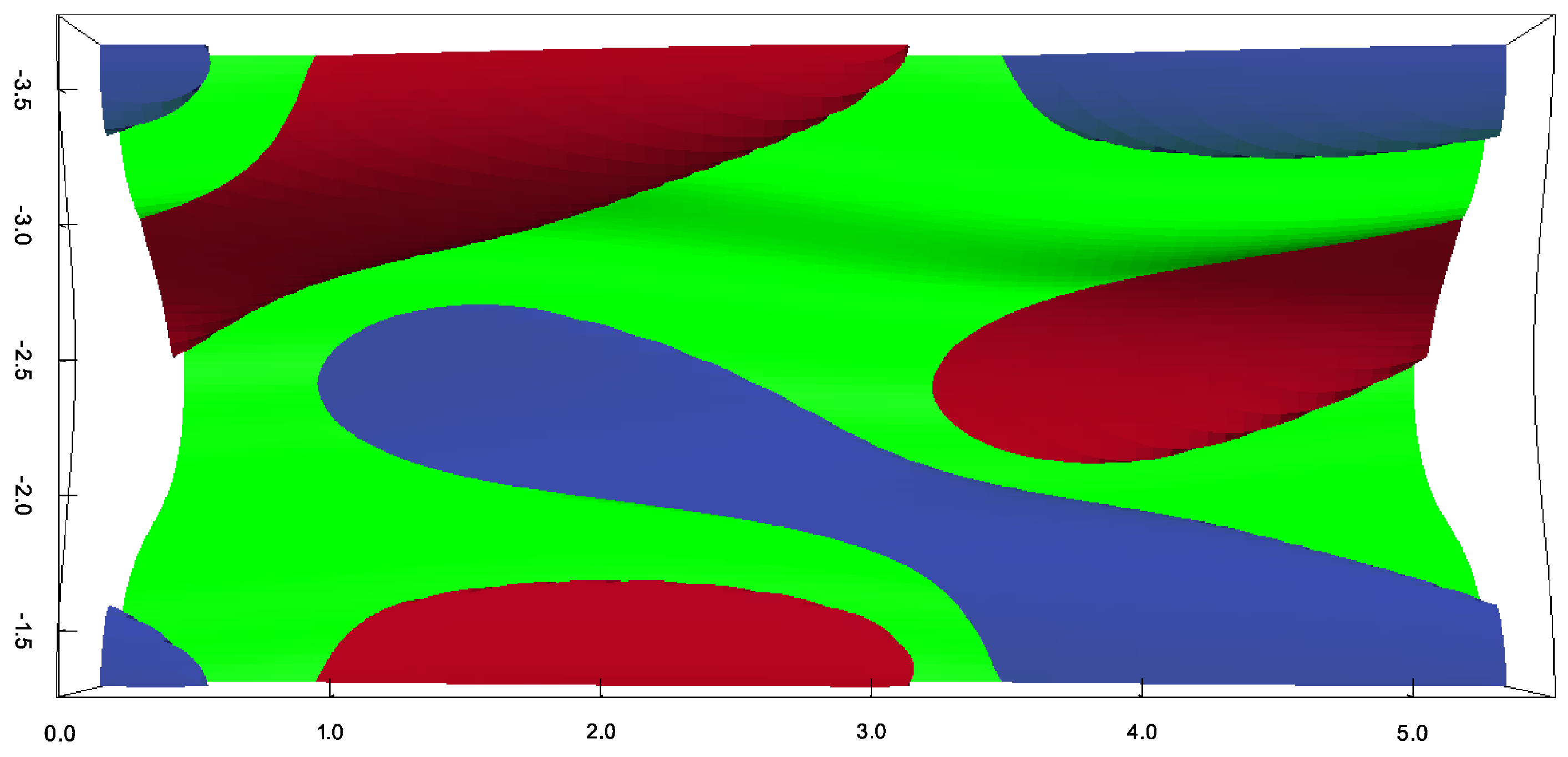}};
			\node[below=of img, node distance=0cm, xshift=-0.2cm,yshift=1.2cm]{$x$};
			\node[left=of img, node distance=0cm, rotate=90, anchor=center, yshift=-0.9cm]{$z$};
		\end{tikzpicture}
		\caption{EQ1 (top view)}
	\end{subfigure}
	\begin{subfigure}{0.44\textwidth}
	\centering
		\begin{tikzpicture}
			\node (img) {\includegraphics[width=0.9\textwidth]{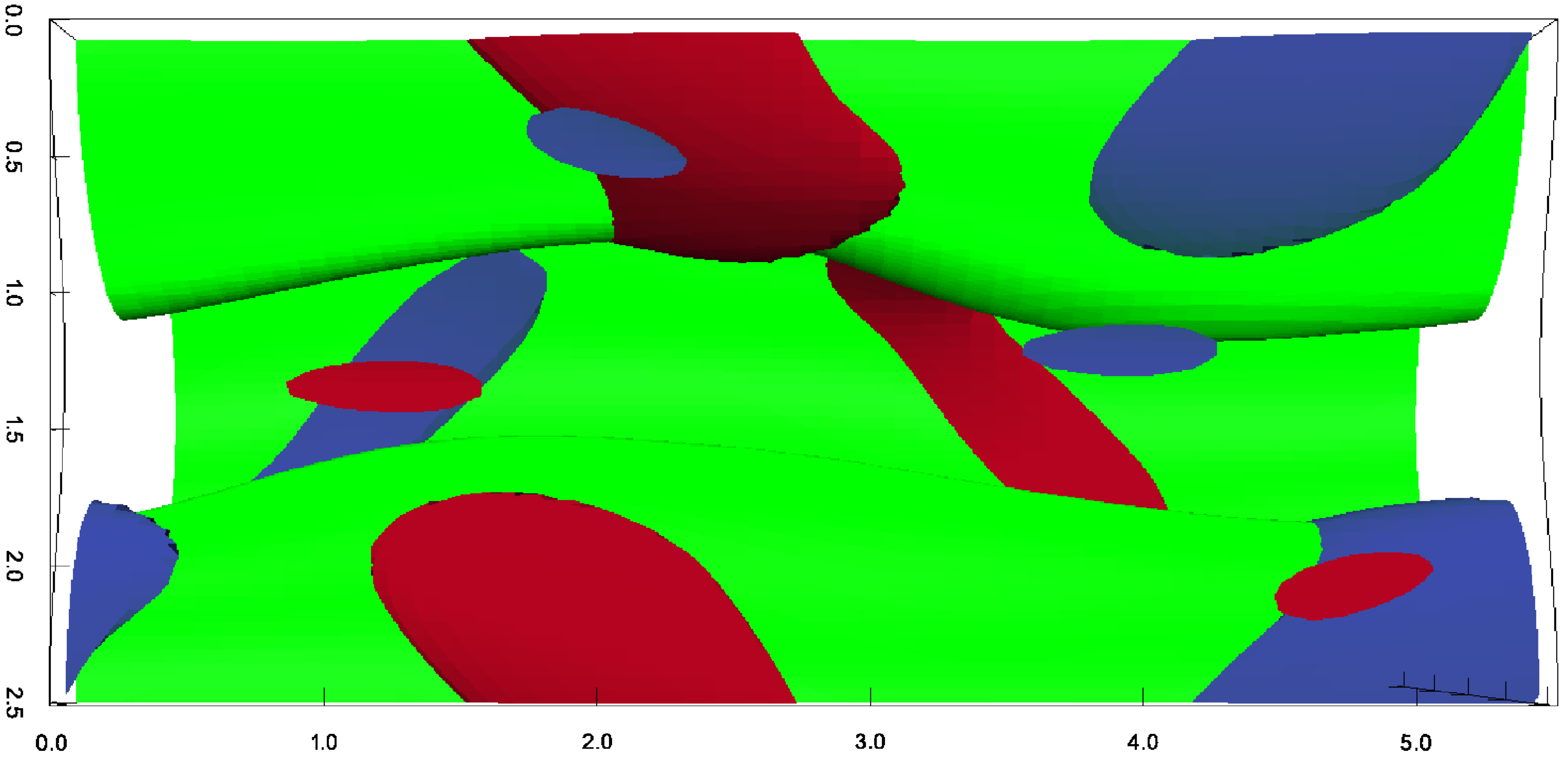}};
			\node[below=of img, node distance=0cm, xshift=-0.2cm,yshift=1.2cm]{$x$};
			\node[left=of img, node distance=0cm, rotate=90, anchor=center, yshift=-0.9cm]{$z$};
		\end{tikzpicture}
		\caption{EQ2 (top view)}
	\end{subfigure}\\
	\begin{subfigure}{0.44\textwidth}
	\centering
		\begin{tikzpicture}
			\node (img) {\includegraphics[width=0.9\textwidth]{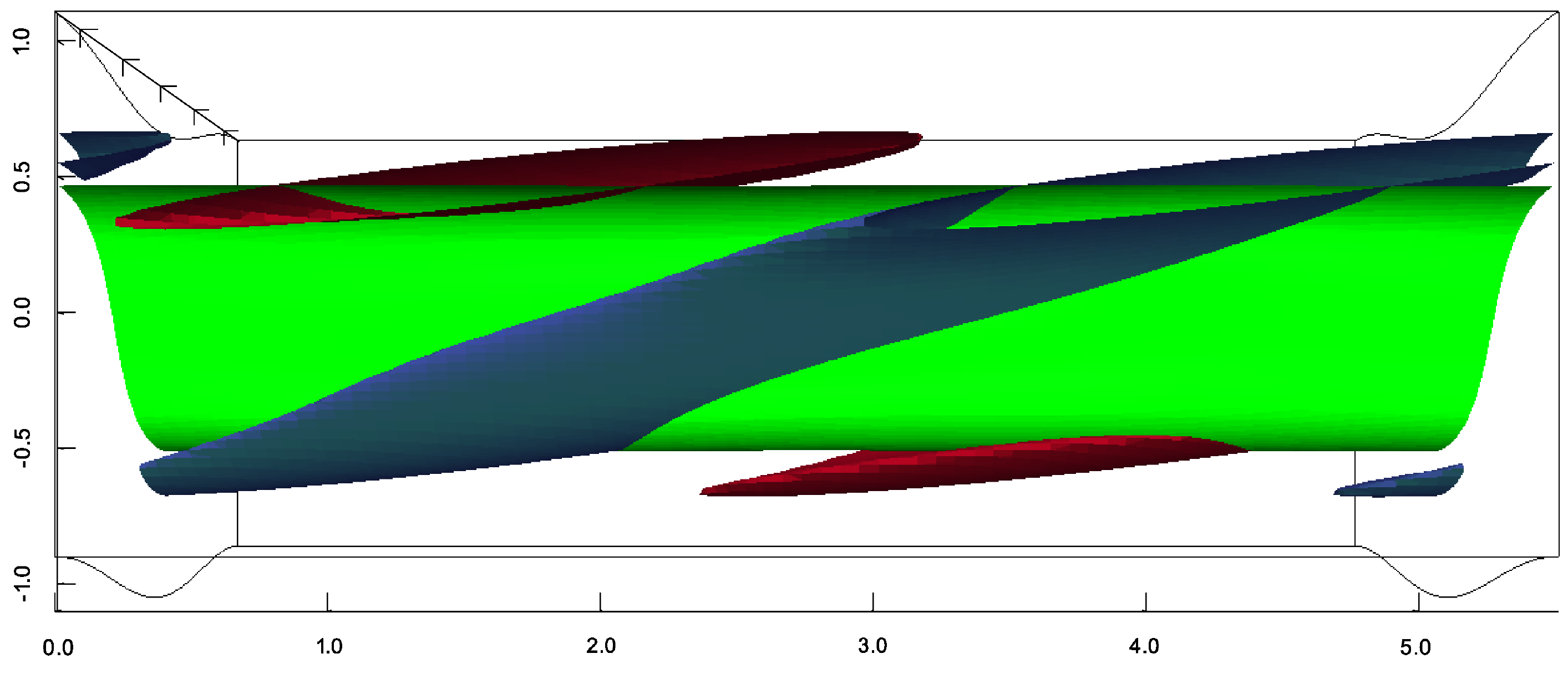}};
			\node[below=of img, node distance=0cm, xshift=-0.2cm,yshift=1.2cm]{$x$};
			\node[left=of img, node distance=0cm, rotate=90, anchor=center, yshift=-0.9cm]{$y$};
		\end{tikzpicture}
		\caption{EQ1 (side view)}
	\end{subfigure}
	\begin{subfigure}{0.44\textwidth}
	\centering
		\begin{tikzpicture}
			\node (img) {\includegraphics[width=0.9\textwidth]{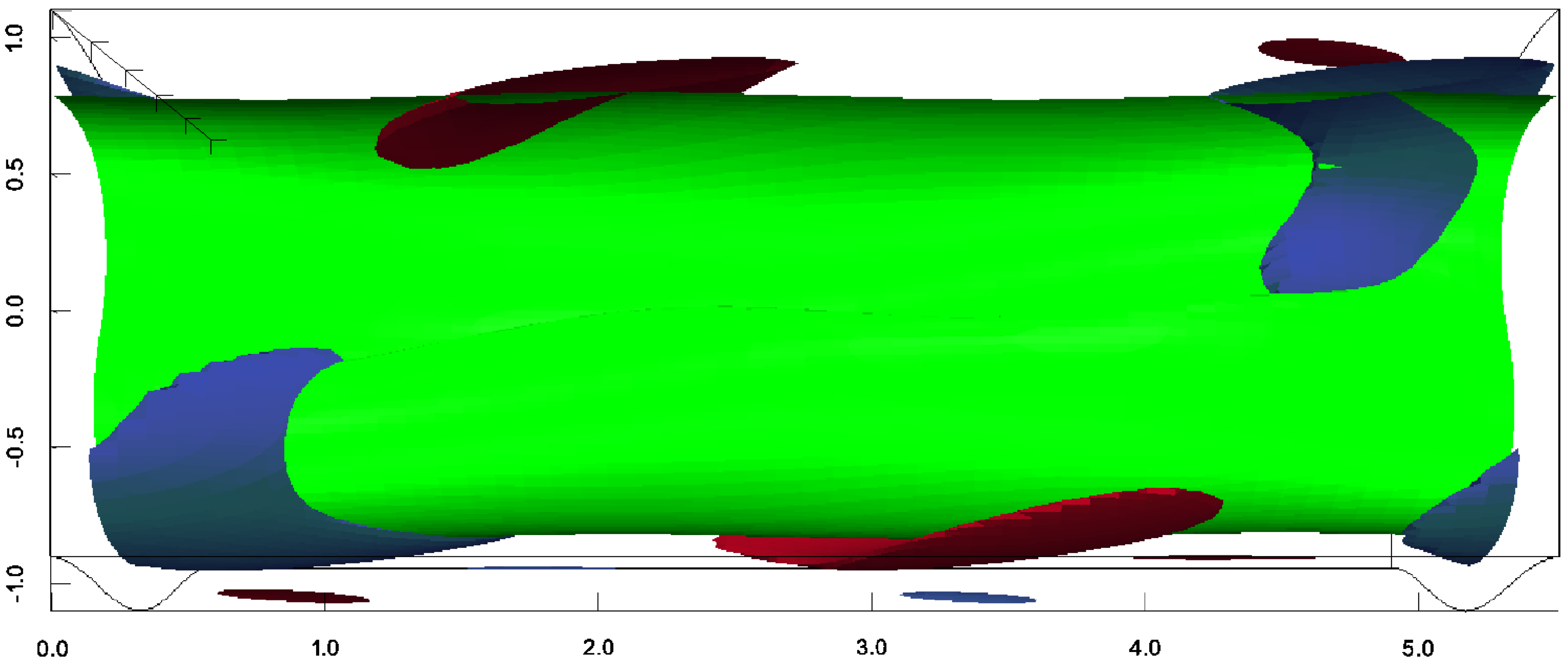}};
			\node[below=of img, node distance=0cm, xshift=-0.2cm,yshift=1.2cm]{$x$};
			\node[left=of img, node distance=0cm, rotate=90, anchor=center, yshift=-0.9cm]{$y$};
		\end{tikzpicture}
		\caption{EQ2 (side view)}
	\end{subfigure}\\
	\caption{Lower (EQ1) and upper (EQ2) branch equilibria for grooved PCF at $A=0.1$ showing isosurfaces for zero streamwise velocity (green) and streamwise vortices for $\omega_x$ = $0.7 \textrm{max}(\omega_{x})$ (red) and $-0.7 \textrm{max}(\omega_{x})$ (blue). \label{fig:3dvelocity}}
\end{figure}

The wavelength of the grooves, defined by $(L_z=2\pi/\beta)$, is set to coincide with the spanwise periodic length of the original flat PCF equilibria: $\beta = 2.5$. The streamwise length of the periodic box is also set according to the equilibria to be continued: $\alpha = 1.14$. All solutions are for $Re=400$. The flat PCF solutions EQ1 (lower branch) and EQ2 (upper branch) are shown in \cref{fig:lower-flat,fig:upper-flat}; contours of streamwise velocity and quiver arrows for cross-stream velocities are plotted. The continued solutions for grooved PCF are shown in \cref{fig:lower-grooved,fig:upper-grooved} at $A_1 = 10\%$. The vortex-streak interaction in grooved PCF can be seen from the isosurfaces of zero streamwise velocity and $\pm 0.7\textrm{max}(\omega_x)$ streamwise vorticity shown in \cref{fig:3dvelocity}. The residual norm for these solutions is lower than $10^{-12}$ and their spatial accuracy is $\sim 10^{-5}$.

Solutions for grooved PCF look similar to their flat PCF counterparts, except for some distortion in the streamwise vortices. Streamwise streaks are accompanied by streamwise vortices centered on the low speed streaks. This is not surprising; the interaction of streamwise streaks and vortices is known to be a hallmark of wall-turbulence and has been shown to exist irrespective of the kind of wall-bounded flow \citep{waleffe1997self,blackburn2013lower}.

\subsection{Phase restriction}\label{sec:localization}
The flowfield shown in \cref{fig:lower-flat} is just one member of a continuous family of solutions that arise due to spanwise homogeneity of flat PCF. This solution satisfies the shift-reflect symmetry $s_1$ about the plane $z=0$. We refer to this solution as the zero-phase solution and denote it $\chi^*$. Let us denote a flowfield that is obtained by translating $\chi^*$ along $z$ by some shift $l_z$ as $S_{l_z \neq 0}(\chi^*)$. This flowfield still satisfies a shift-reflect symmetry, but about the plane $z=l_z$ instead of $z=0$; we say that these solutions have non-zero phases, with the phase being $\tan^{-1}(l_z/L_z) $. The Fourier coefficients, denoted $\mathscr{F}\{ \cdot \}_{k,m}$ for mode $e^{i(k\alpha X +  m\beta Z)}$, of $S_{l_z}(\chi^*)$ can be related to the Fourier coefficients of $\chi^*$ as 
\begin{equation}
	\mathscr{F}\{S_{l_z}(\chi^*)\}_{k,m}(Y) = e^{-2\pi i m(l_z/L_z)}\mathscr{F}\{\chi^*\}_{k,m}(Y),
\end{equation}
so that spanwise translation of flowfields can be seen as phase-shifts in the Fourier coefficients.  

Let us also define the continuous family of flowfields obtained by translating $\chi^*$ along $z$ as $\tilde{\mathcal{F}}(\chi^*) = \{ S_{\delta z}(\chi^*): \delta z \in \mathbb{R}\}$, and the discrete family of flowfields obtained by translating $\chi^*$ along $z$ through integral multiples of $L_z/n$ as $\mathcal{F}_n(\chi^*) = \{S_{q L_z/n}(\chi^*) : q \in \mathbb{Z}\}$. The argument $\chi^*$ for $\mathcal{F}$ is dropped for convenience. Every $\mathcal{F}_n$ is a subset of $\tilde{\mathcal{F}}$. The members of $\mathcal{F}_1$ are identical to each other due to the periodicity of $\chi^*$. 

For flat PCF, because of its spanwise homogeneity, every member of $\tilde{\mathcal{F}}$ is an equilibrium of the NSE. When $\chi^*$ is used as the initial iterate for finding a grooved PCF equilibrium with groove amplitude $A$, it converges to $\chi^*_A$ (shown in \cref{fig:3dvelocity} for $A=0.1$). However, when a different member of $\tilde{\mathcal{F}}$ is used as the initial iterate, convergence is not likely. We find that the iterations converge only for flowfields $\chi \in \mathcal{F}_2$, or for flowfields sufficiently close to members of $\mathcal{F}_2$; the converged solutions for the latter case are identical to the solutions obtained from continuing the corresponding member in $\mathcal{F}_2$. Failure of iterations to converge indicates that the initial iterate is not in the vicinity of an equilibrium. We take this as indicating that most members of $\tilde{\mathcal{F}}$ do not have a corresponding equilibrium in grooved PCF. 

The significance of $\mathcal{F}_2$ can be attributed to a discrete symmetry of PCF. The original equilibria, EQ1 and EQ2, are invariant under the discrete symmetries $s_1$, $s_2$, and $s_3$.
The symmetry $s_1$ is a composition of reflection along the $x-y$ plane $z=0$, and a half-cell shift along $x$; the reflection and $x$-shift commute. Because of the spanwise homogeneity of flat PCF, it admits solutions with reflectional symmetry about any $x-y$ plane, such as a flowfield $\chi = S_{l_z}(\chi^*)$ which is shift-reflect invariant about the plane $z=l_z$. By contrast, grooved PCF has reflectional symmetry only about particular $x-y$ planes, as can be seen in \cref{fig:groovedPCF}. In $\lbrack 0,2\pi/\beta)$, reflectional symmetry holds only for planes $z=0$ and $z=\pi/\beta$. So, any equilibrium of grooved PCF can have reflectional symmetry only about these two planes. Initial iterates from $\mathcal{F}_2$ conform to this symmetry even if they do not exactly satisfy the NSE, while flowfields in $\tilde{\mathcal{F}} - \mathcal{F}_2$ do not even satisfy the symmetry. Our results show that the solutions obtained by continuing $\mathcal{F}_2$ also satisfy the shift-reflect symmetry $s_1$. 

\subsubsection{Bifurcation}
$\mathcal{F}_2$ has two non-identical members: $\chi^*$ and $\chi^+ = S_{L_z/2}(\chi^*)$. Let us denote their continued solutions at some amplitude $A$ by $\chi^*_A$ and $\chi^+_A$ respectively. Our results show that $\chi^+_A \neq S_{L_z/2}(\chi^*_A)$; the members of $\mathcal{F}_2(\chi^*)$ bifurcate into two distinct families of solutions in grooved PCF: $\mathcal{F}_1(\chi^*_A)$ and $\mathcal{F}_1(\chi^+_A)$, with distinct values for scalars such as energy density and bulk dissipation rate. The difference arises because, while grooved PCF has reflectional symmetry about both $z=0$ and $z=\pi/\beta$, the walls at $z=0$ are displaced along $+y$ and the walls at $z=\pi/\beta$ are displaced along $-y$. This bifurcation can also be interpreted in terms of continuing $\chi^*$ along positive and negative groove amplitudes.

\subsubsection{Spatial anchoring}
\Citet{willis2013revealing} argue that the presence of continuous families of solutions could obscure our description of turbulent flow and describe a method of slices to systematically reduce a family of similar solutions to one representative solution; with this approach, travelling wave solutions reduce to equilibria, and relative periodic orbits to periodic orbits. Grooved PCF achieves this reduction by physically disrupting some of these symmetries. Spatially localized equilibria have also been found in flat PCF \citep[for instance ][]{gibson2014spanwise,chantry2015localization}. Based on our results for global solutions, we can speculate that any symmetric localized solutions, when extended to grooved PCF, would also be phase-restricted by the grooves. 

In flat PCF turbulence, any of the solutions in a family $\tilde{\mathcal{F}}$ could be visited; no member of a family is special. As a result, turbulence statistics would only show spatial averages over an entire family of solutions. The phase-restriction of solutions in grooved PCF, however, implies that turbulence statistics should reflect the spatial structure of the solutions. 

\begin{figure}
	\centering
	\begin{subfigure}{0.47\textwidth}
		\begin{tikzpicture}
			\node (img) {\includegraphics[width=\textwidth]{velocityE1_A00x00.pdf}};
			\node[below=of img, node distance=0cm, xshift=-0.2cm,yshift=1.2cm]{$z$};
			\node[left=of img, node distance=0cm, rotate=90, anchor=center, yshift=-1.1cm]{$y$};
		\end{tikzpicture}
		\caption{Smooth PCF}
	\end{subfigure}
	\begin{subfigure}{0.47\textwidth}
		\begin{tikzpicture}
			\node (img) {\includegraphics[width=\textwidth]{velocityE1_A10x00.pdf}};
			\node[below=of img, node distance=0cm, xshift=-0.2cm,yshift=1.2cm]{$z$};
			\node[left=of img, node distance=0cm, rotate=90, anchor=center, yshift=-1.1cm]{$y$};
		\end{tikzpicture}
		\caption{Grooved PCF with $A_1 = 10\%$}
	\end{subfigure}\\
	\begin{subfigure}{0.47\textwidth}
		\begin{tikzpicture}
			\node (img) {\includegraphics[width=\textwidth]{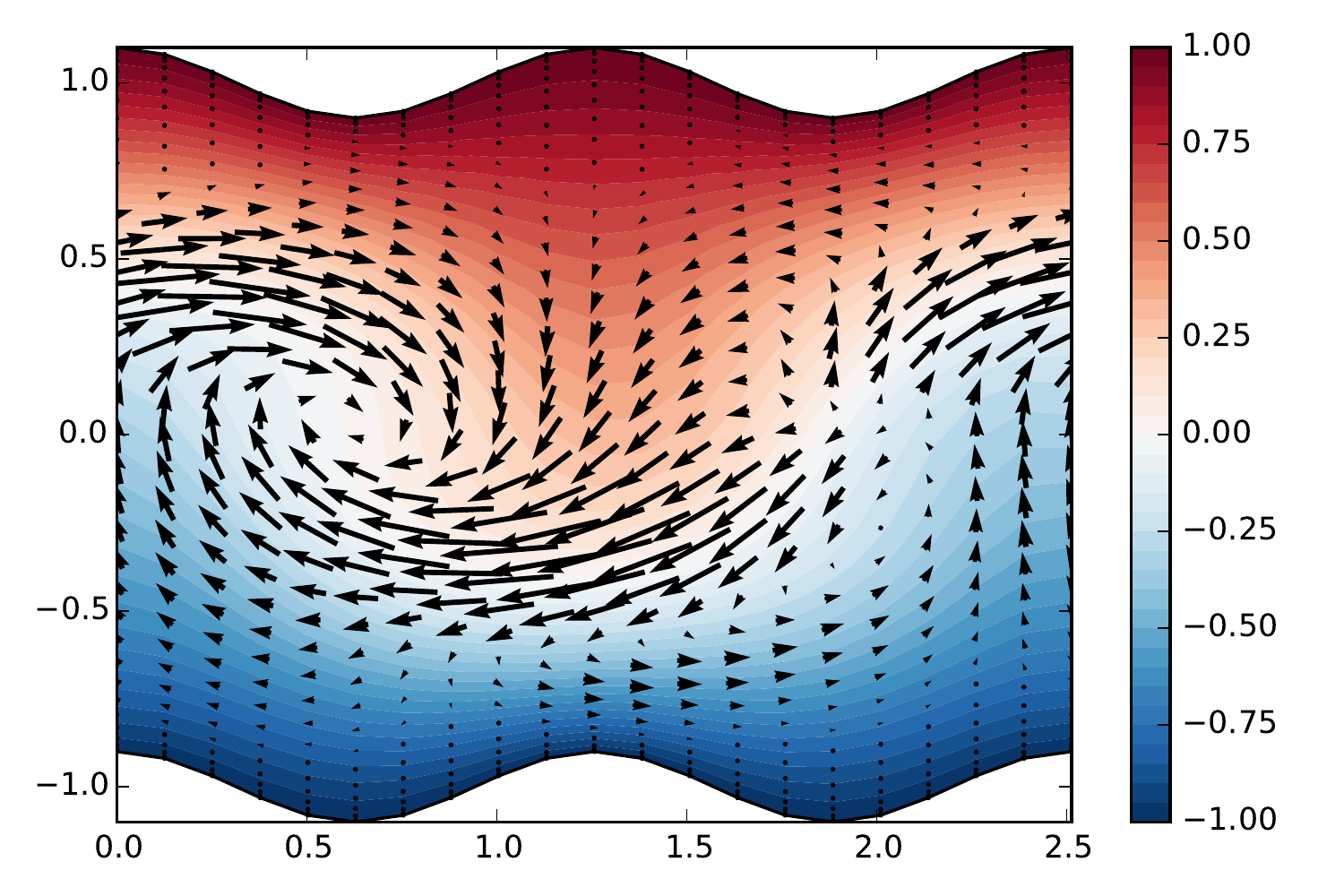}};
			\node[below=of img, node distance=0cm, xshift=-0.2cm,yshift=1.2cm]{$z$};
			\node[left=of img, node distance=0cm, rotate=90, anchor=center, yshift=-1.1cm]{$y$};
		\end{tikzpicture}
		\caption{Grooved PCF with $A_2=10\%$}
	\end{subfigure}
	\begin{subfigure}{0.47\textwidth}
		\begin{tikzpicture}
			\node (img) {\includegraphics[width=\textwidth]{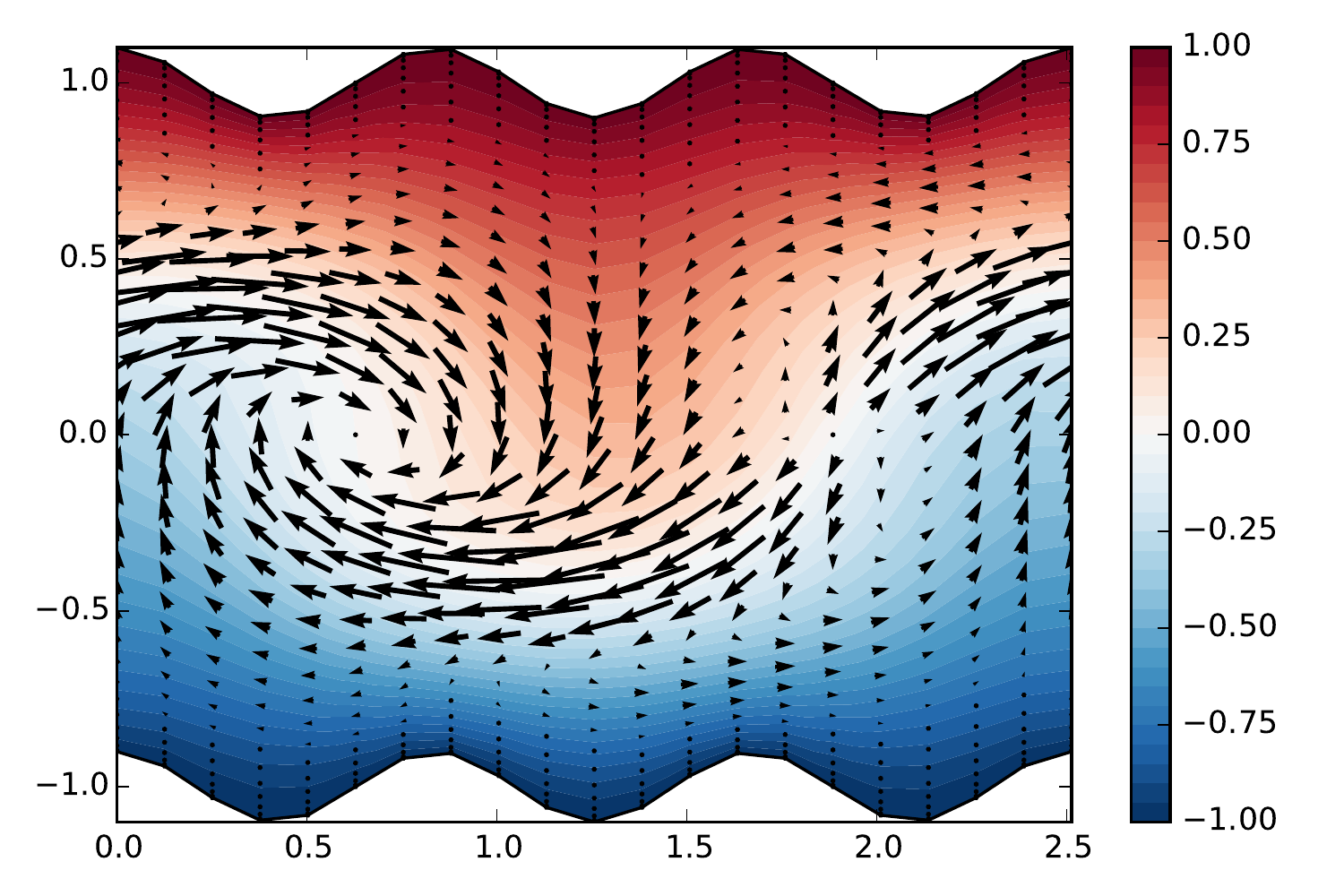}};
			\node[below=of img, node distance=0cm, xshift=-0.2cm,yshift=1.2cm]{$z$};
			\node[left=of img, node distance=0cm, rotate=90, anchor=center, yshift=-1.1cm]{$y$};
		\end{tikzpicture}
		\caption{Grooved PCF with $A_3 = 10\%$}
	\end{subfigure}
	\caption[EQ1 in grooved PCF with one, two, and three grooves per box]{Velocity on a cross-stream plane ($x=0$) for flat PCF (a) and grooved PCF for one, two, and three grooves per box (b-d) for the lower branch solution EQ1. Contours show streamwise velocity and quiver arrows show cross-stream velocities.\label{fig:velocity-E1-3}}
\end{figure}

\begin{figure}
	\centering
	\begin{subfigure}{0.47\textwidth}
		\begin{tikzpicture}
			\node (img) {\includegraphics[width=\textwidth]{velocityEQ2E1_A00x00.pdf}};
			\node[below=of img, node distance=0cm, xshift=-0.2cm,yshift=1.2cm]{$z$};
			\node[left=of img, node distance=0cm, rotate=90, anchor=center, yshift=-1.1cm]{$y$};
		\end{tikzpicture}
		\caption{Smooth PCF}
	\end{subfigure}
	\begin{subfigure}{0.47\textwidth}
		\begin{tikzpicture}
			\node (img) {\includegraphics[width=\textwidth]{velocityEQ2E1_A10x00.pdf}};
			\node[below=of img, node distance=0cm, xshift=-0.2cm,yshift=1.2cm]{$z$};
			\node[left=of img, node distance=0cm, rotate=90, anchor=center, yshift=-1.1cm]{$y$};
		\end{tikzpicture}
		\caption{Grooved PCF with $A_1 = 10\%$}
	\end{subfigure}\\
	\begin{subfigure}{0.47\textwidth}
		\begin{tikzpicture}
			\node (img) {\includegraphics[width=\textwidth]{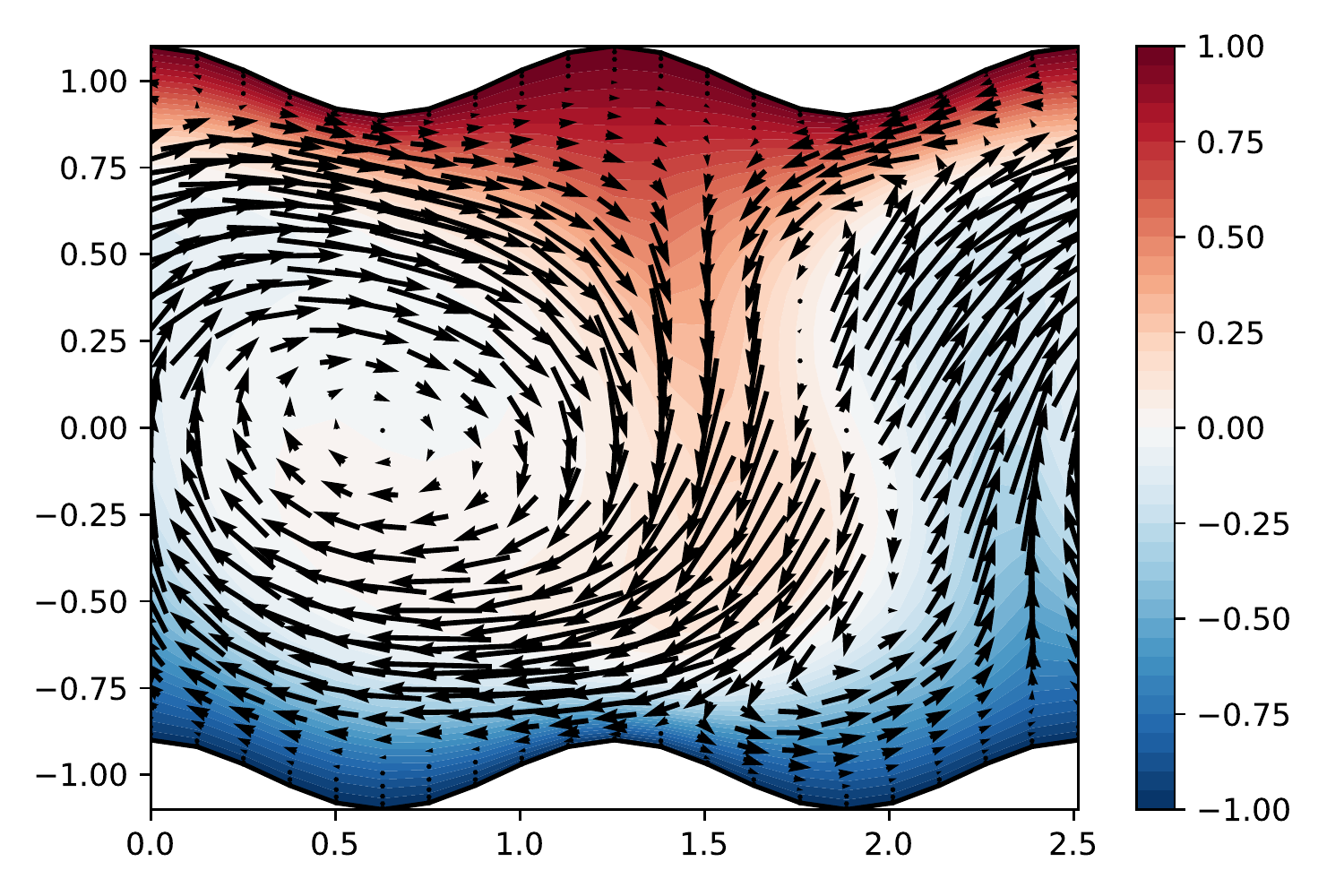}};
			\node[below=of img, node distance=0cm, xshift=-0.2cm,yshift=1.2cm]{$z$};
			\node[left=of img, node distance=0cm, rotate=90, anchor=center, yshift=-1.1cm]{$y$};
		\end{tikzpicture}
		\caption{Grooved PCF with $A_2=10\%$}
	\end{subfigure}
	\begin{subfigure}{0.47\textwidth}
		\begin{tikzpicture}
			\node (img) {\includegraphics[width=\textwidth]{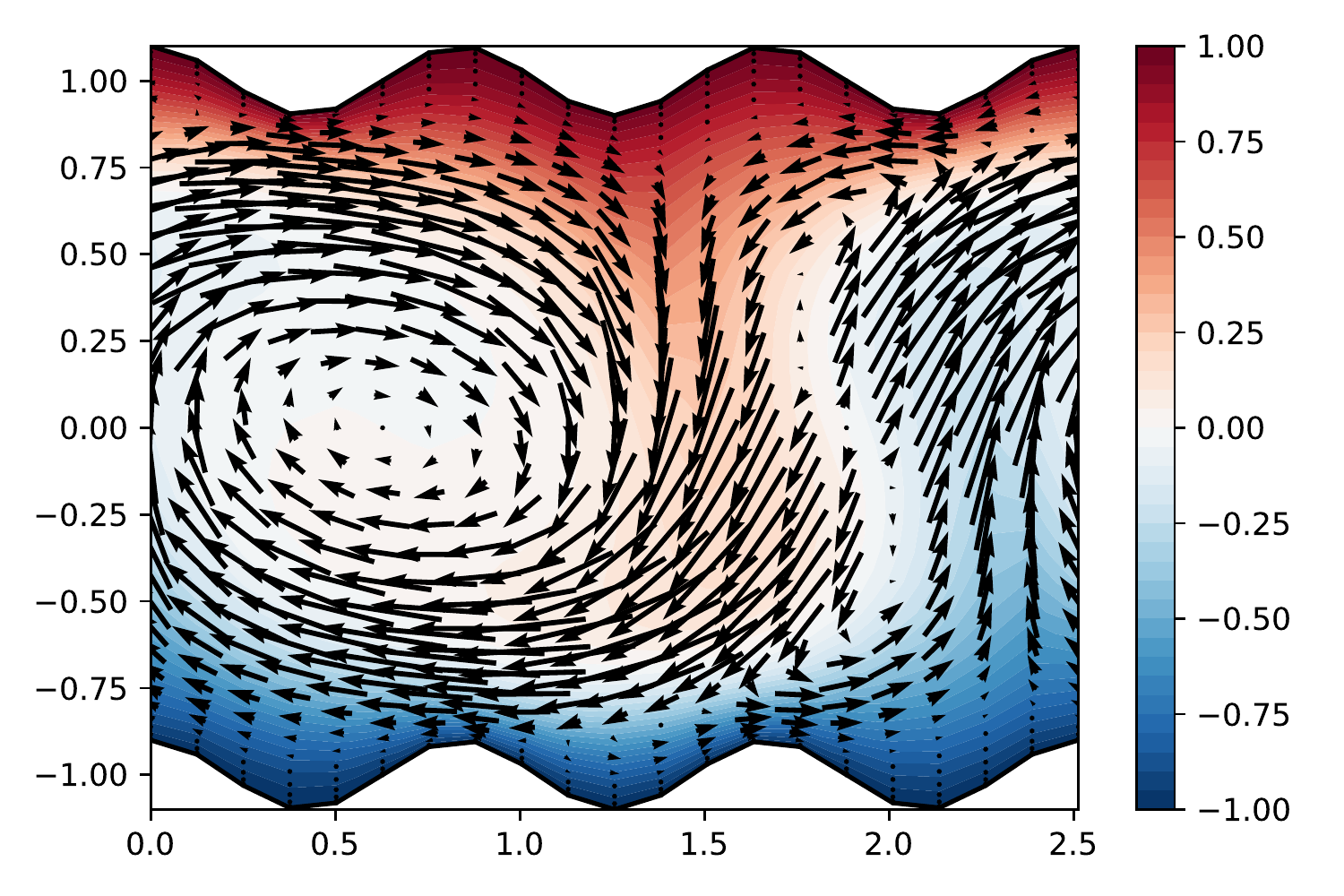}};
			\node[below=of img, node distance=0cm, xshift=-0.2cm,yshift=1.2cm]{$z$};
			\node[left=of img, node distance=0cm, rotate=90, anchor=center, yshift=-1.1cm]{$y$};
		\end{tikzpicture}
		\caption{Grooved PCF with $A_3 = 10\%$}
	\end{subfigure}
	\caption[EQ2 in grooved PCF with one, two, and three grooves per box]{Velocity on a cross-stream plane ($x=0$) for flat PCF (a) and grooved PCF for one, two, and three grooves per box (b-d) for the upper branch solution EQ2. Contours show streamwise velocity and quiver arrows show cross-stream velocities.\label{fig:velocity-EQ2-E1-3}}
\end{figure}

\subsection{Wavelength dependence}
We now introduce multiple grooves within the periodic box of size $L_z = 2\pi/\beta = 2\pi/2.5$, with the walls given for these cases by 
\begin{equation}\label{multiple-groove}
    y_{walls} = \pm 1 + \frac{A_k}{5} \cos( \beta z) + A_k \cos(k\beta z).
\end{equation}
The $\cos(\beta z)$ term is included to retain the fundamental periodicity along $z$.

Figure \ref{fig:velocity-E1-3} illustrates continuation of EQ1 to three different geometries --- one, two, and three grooves per box --- at a dominant groove amplitude of $10\%$. EQ2 is plotted for the same cases in \cref{fig:velocity-EQ2-E1-3}. The two groove per box geometry does not satisfy the shift-rotate symmetry $s_2$ or the shift-invert symmetry $s_3$; these cases were run at a resolution of 20x35x32 grid points ($(L,M,N)=(10,16,35)$) to converge to residual norms $\sim 10^{-7}$ and accuracies $\sim 10^{-5}$ for EQ1 and $\sim 10^{-3}$ for EQ2. The three groove per box satisfies $s_1$, $s_2$, and $s_3$, and these cases were run with 28x35x32 grid points ($(L,M,N)=(14,16,35)$) like the one groove per box case, and have residual norms similar to the one groove per box case. 
The vortex-streak structure remains the same for the two and three grooves per box cases as well. However, the large scale distortion seen in the one groove per box case is not as pronounced in the other two cases. 

\subsubsection{Mean velocity}
\begin{figure}
	\centering
	\begin{subfigure}{0.47\textwidth}
		\includegraphics[width=\textwidth]{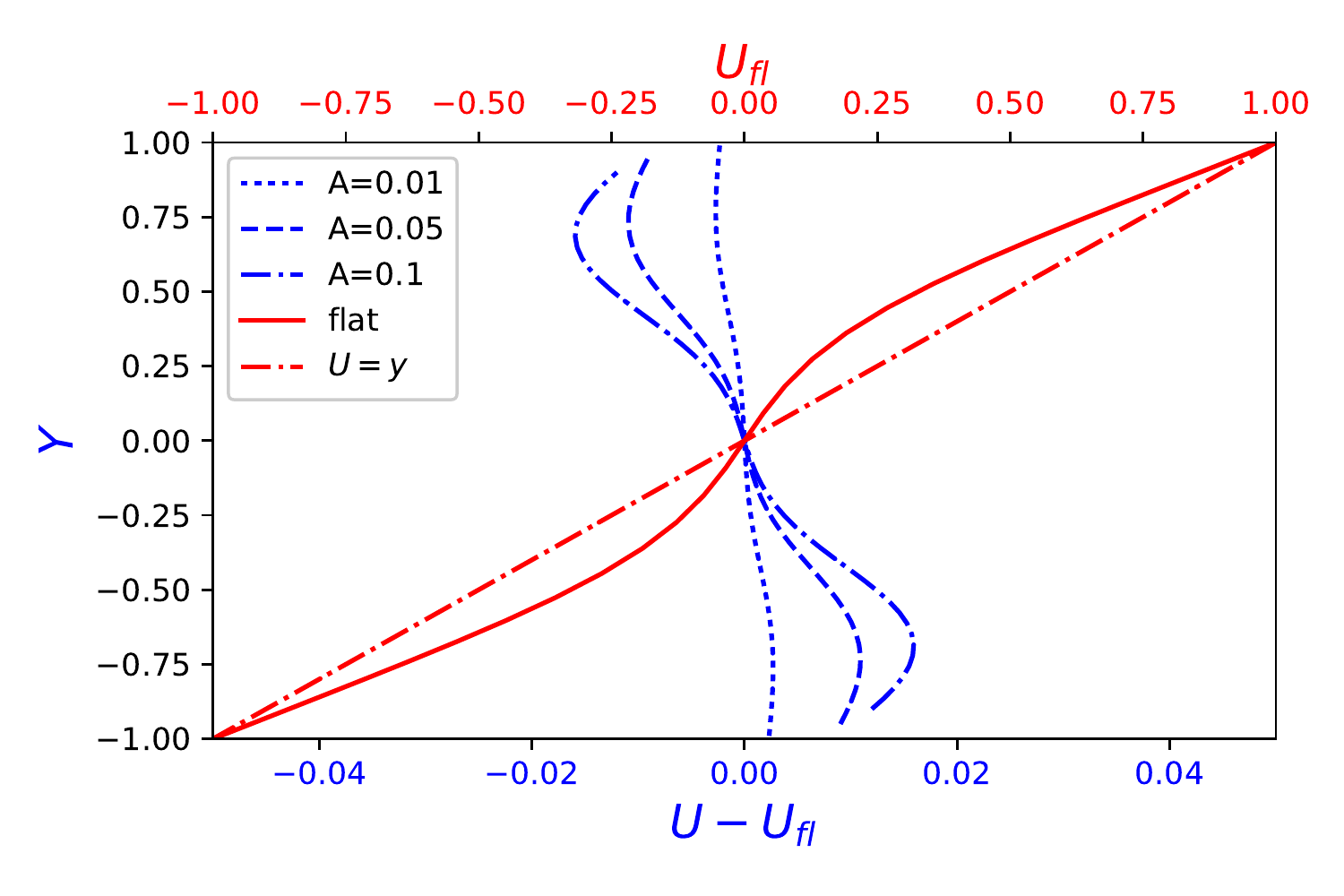}
		\caption{EQ1; one groove per box}
	\end{subfigure}
	\begin{subfigure}{0.47\textwidth}
		\includegraphics[width=\textwidth]{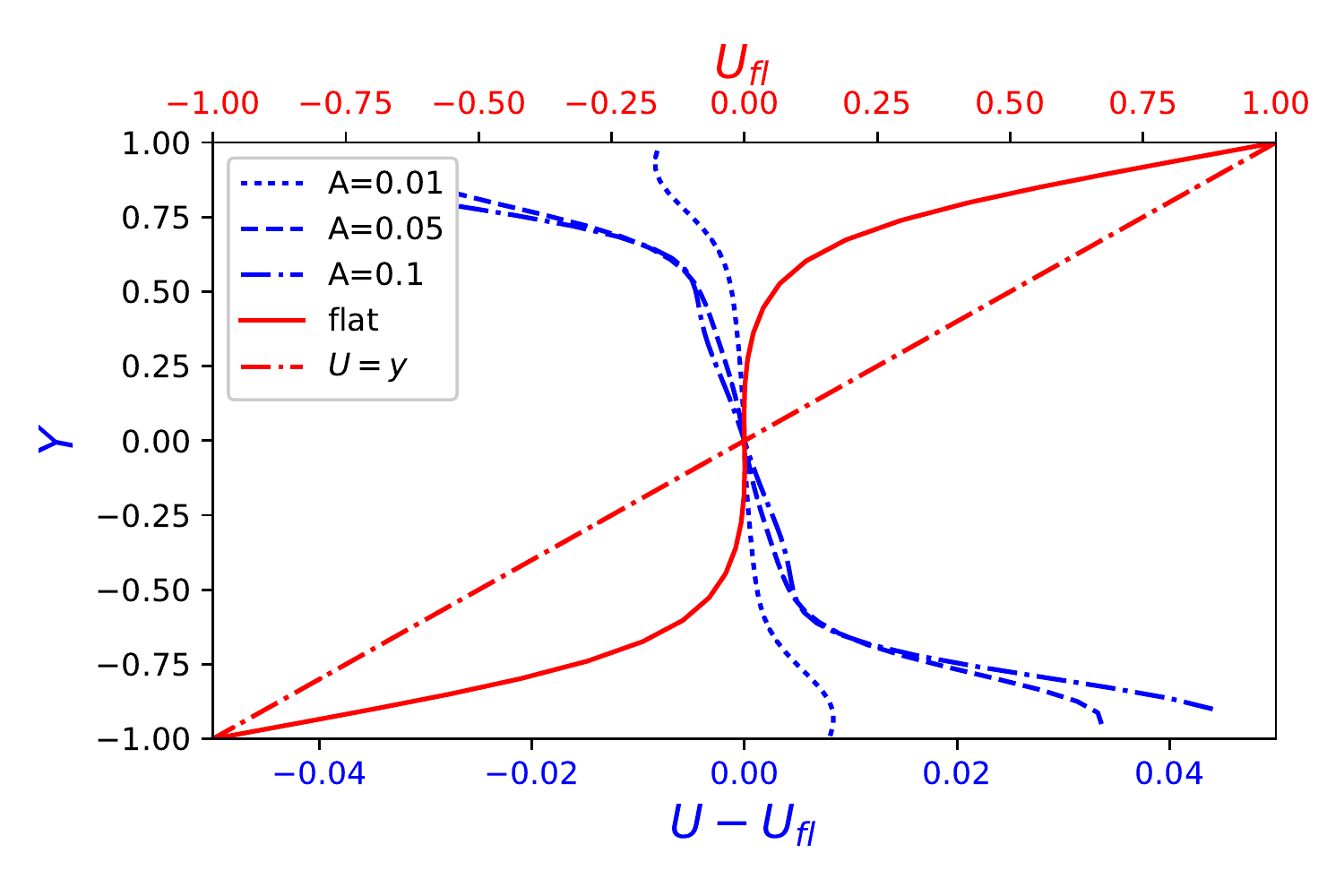}
		\caption{EQ2; one groove per box}
	\end{subfigure}\\
	\begin{subfigure}{0.47\textwidth}
		\includegraphics[width=\textwidth]{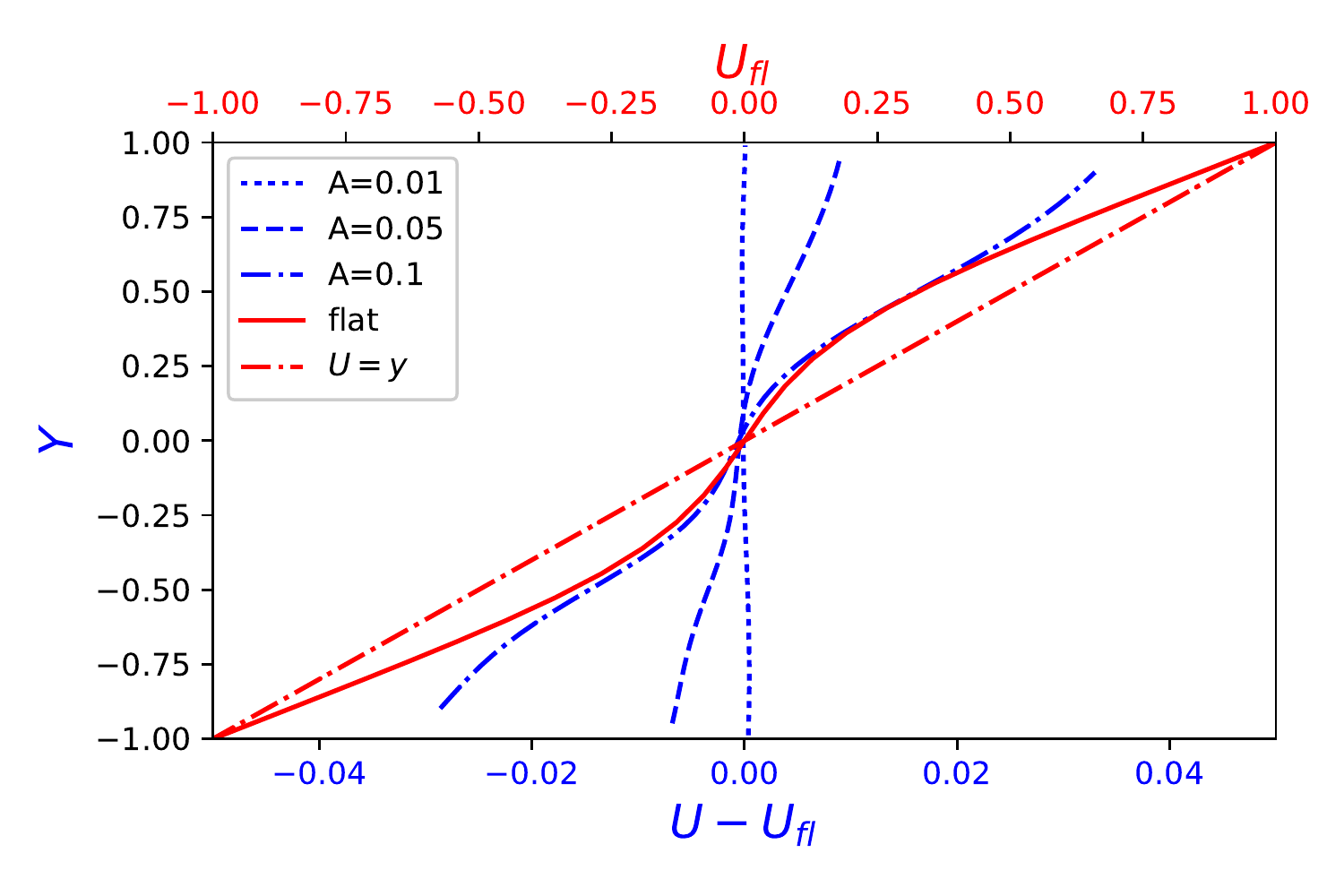}
		\caption{EQ1; two grooves per box}
	\end{subfigure}
	\begin{subfigure}{0.47\textwidth}
		\includegraphics[width=\textwidth]{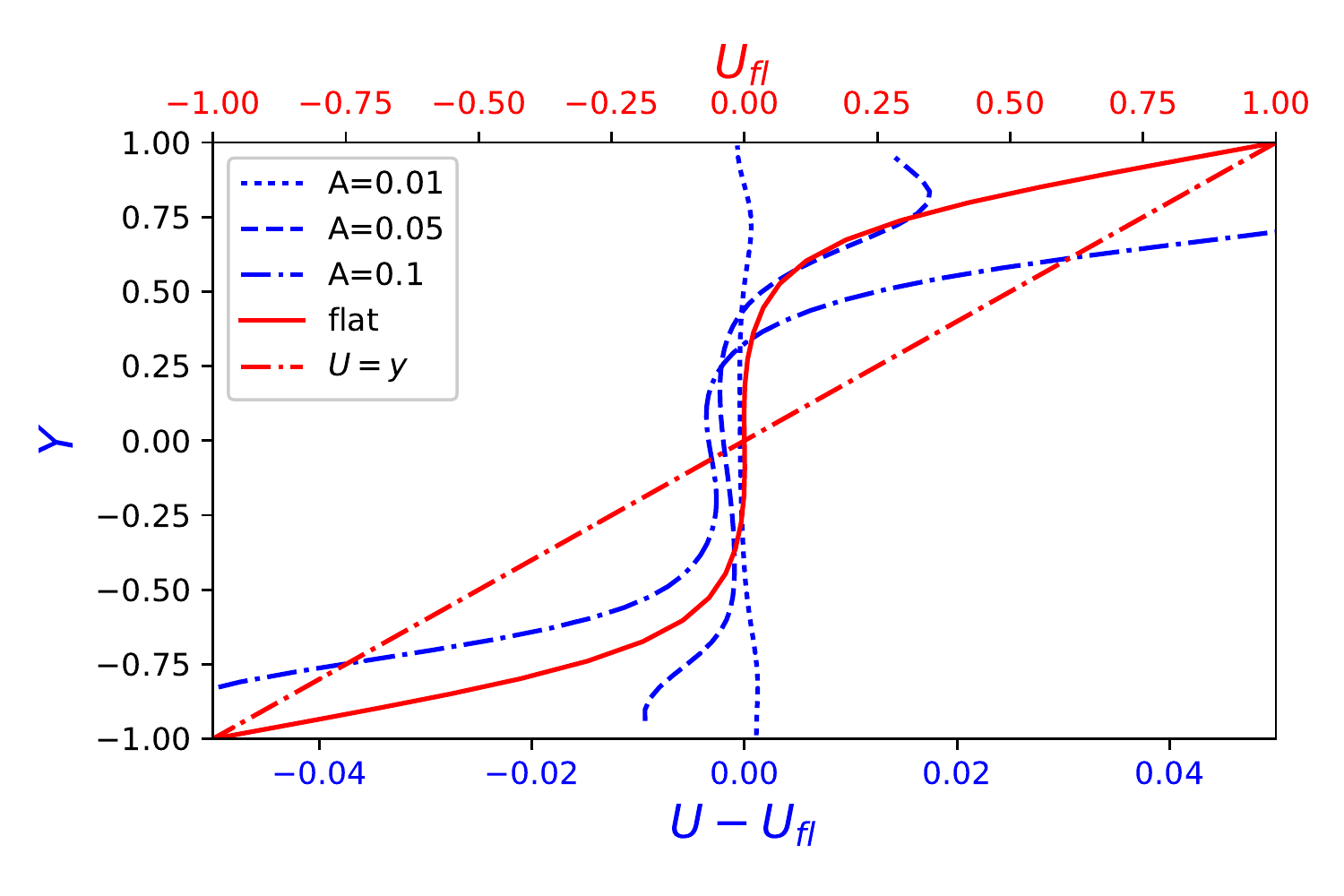}
		\caption{EQ2; two grooves per box}
	\end{subfigure}\\
	\begin{subfigure}{0.47\textwidth}
		\includegraphics[width=\textwidth]{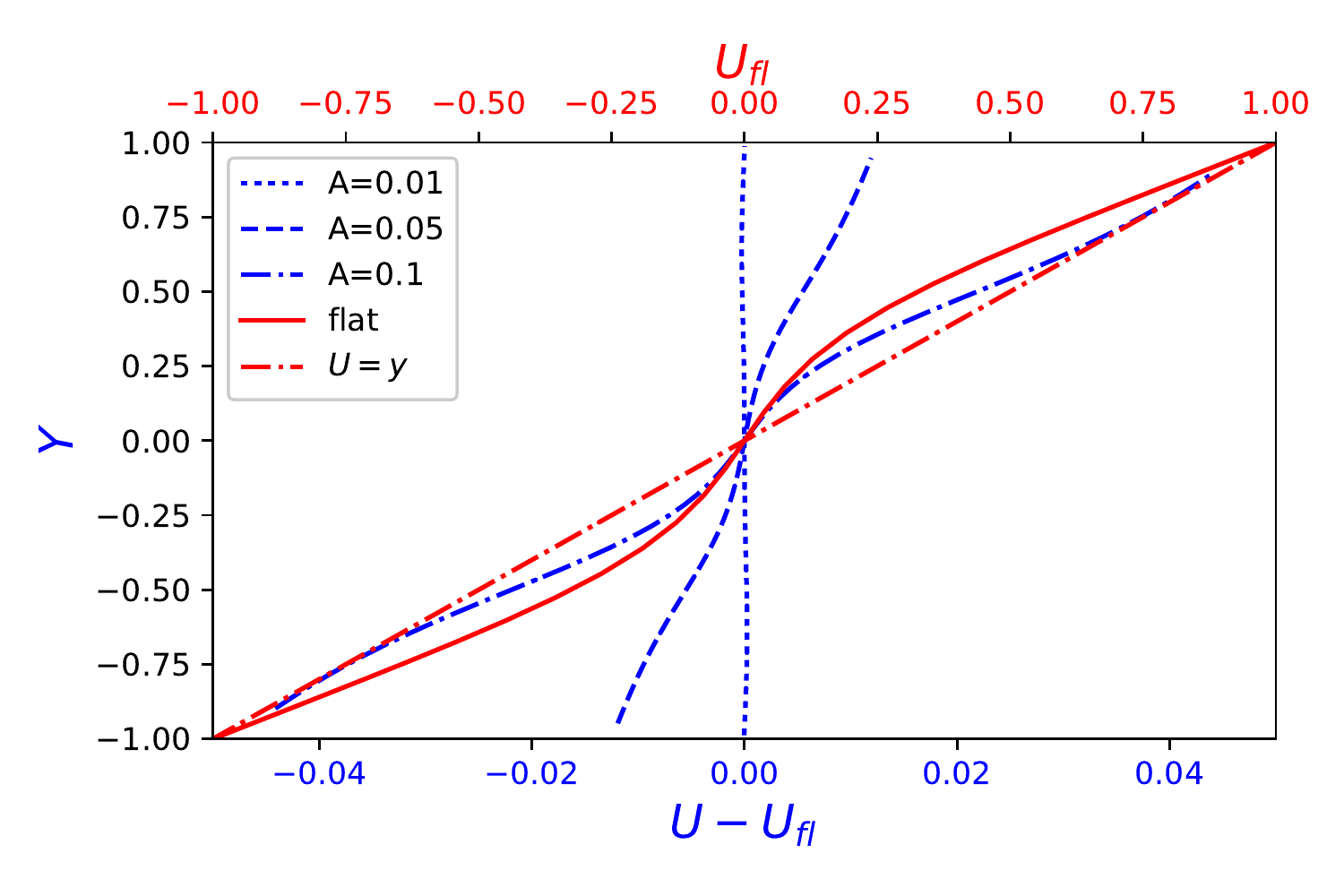}
		\caption{EQ1; three grooves per box}
	\end{subfigure}
	\begin{subfigure}{0.47\textwidth}
		\includegraphics[width=\textwidth]{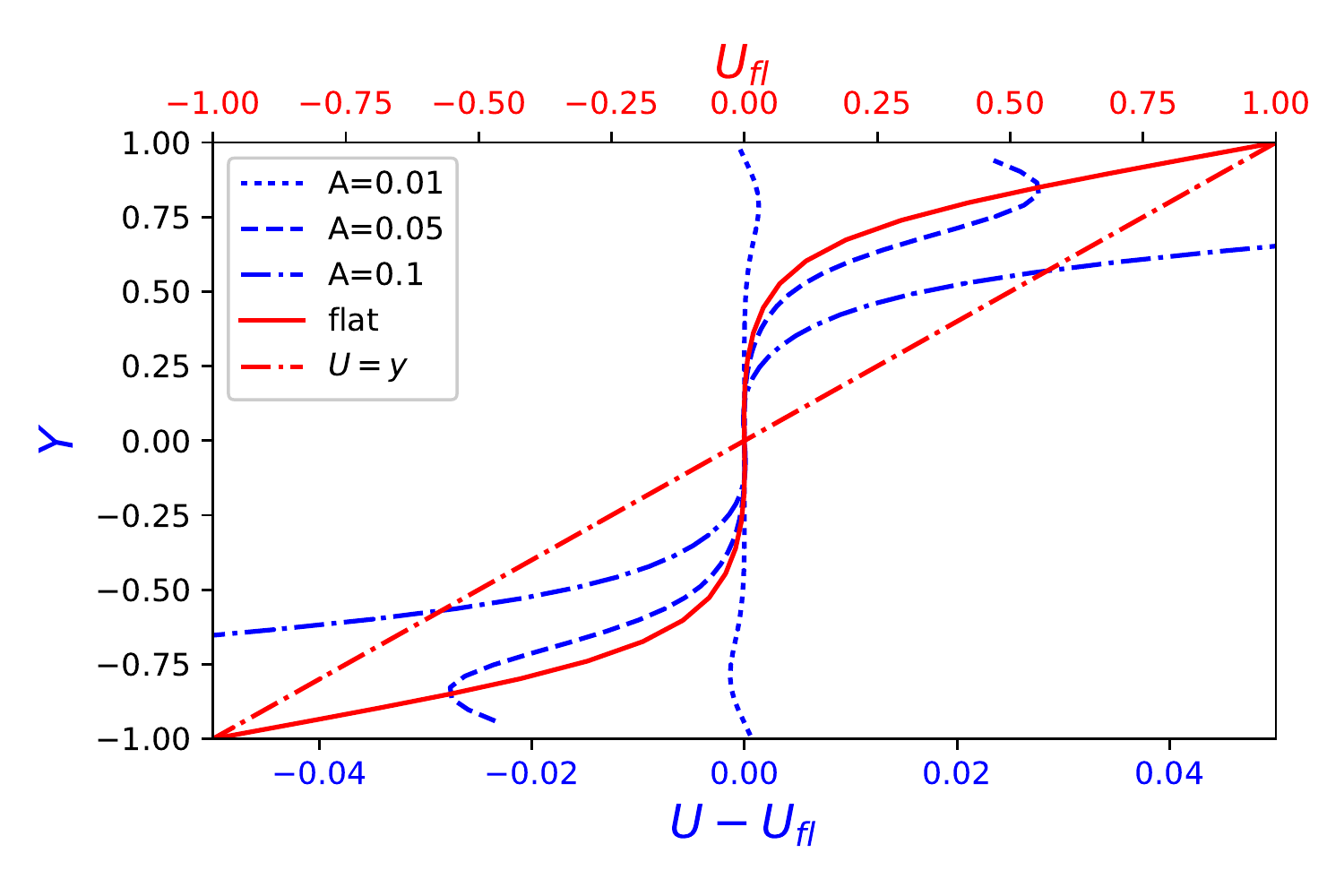}
		\caption{EQ2; three grooves per box}
	\end{subfigure}\\
	\caption[Mean velocity for grooved PCF with multiple grooves; EQ1 and EQ2.]{Mean velocity profiles for EQ1 and EQ2 in grooved PCF with one, two, and three grooves per periodic box, shown as differences from the flat case. The laminar mean velocity, $U=y$, is included for comparison.\label{fig:U-EQ1-EQ2-123} }
\end{figure}

The mean velocity profiles for all three wavenumber cases are shown in \cref{fig:U-EQ1-EQ2-123}, with the profiles for grooved PCF shown as a difference from the mean velocity for flat PCF. In the profiles of the one groove per box cases, the deficit in the mean velocity (compared to $U=y$) is increased by increasing groove-amplitude. In contrast, the profiles of the two and three groove per box cases bridge the velocity deficit in the mean velocity of both EQ1 and EQ2, and the mean velocity grows closer to the $U=y$ profile as the amplitude increases. The difference in mean velocity for EQ2 for the two groove per box is not zero at the centerline; this is because of the loss of the shift-rotate symmetry $s_2$ (see \cref{sec:discrete-symmetries}) for the two groove per box case. 

The bridging of the velocity deficit is stronger for EQ2 than EQ1, although the velocity deficit for EQ2 is much higher than that for EQ1 in the first place. This effect is also stronger for the three grooves per box case than it is for the two grooves per box case. This decrease in velocity deficit also leads to the shear stress at the wall for grooved PCF becoming closer to that for the laminar flow in flat PCF. The one groove per box case increases the shear stress at the wall compared to the flat wall case, while the two and three groves per box cases reduce the shear stress at the wall. While this hints at a drag reducing tendency for the two and three grooves per box cases, the profiles plotted in \cref{fig:U-EQ1-EQ2-123} extend only until the tips of the grooves. They do not show the actual stresses at the wall. An exact quantification of the drag is provided next through plots of dissipation rate. 

\subsubsection{Energy density and dissipation rate}
The kinetic energy density $E$, bulk dissipation rate $D$, and power input $I$ of a velocity field of plane Couette flow are given as
\begin{align}
        E(t) &= \frac{1}{\vol} \int_{\Omega} \frac{1}{2} | \mathbf{u} |^2 d\mathbf{x}, \nonumber\\
        D(t) &= \frac{1}{\vol} \int_{\Omega} |\mathbf{\nabla}\times \mathbf{u} |^2 d\mathbf{x}, \\
        I(t) &= \frac{1}{2\mathcal{A}} \int_{\mathcal{A}} \bigg(\frac{\partial u}{\partial y}\bigg|_{y=1} + \frac{\partial u}{\partial y}\bigg|_{y=-1} \bigg) dx d\zeta,\nonumber
\end{align}
where $\vol = 2L_x L_z$ is the volume of the periodic domain, $d\zeta$ is an infinitesimal arc-length along the wall in the spanwise direction, and $\mathcal{A}$ is the wetted surface area of each wall. These quantities are normalized so that $E=D = 1$ for laminar flow in flat PCF, and $\dot{E} = I-D$. For equilibria, $E$, $I$, and $D$ are not time-dependent, and $I=D$. 

Following the equality of power input and bulk dissipation rate, the drag coefficient ($C_D$) can be related to the bulk dissipation rate ($D$) as
\begin{equation}
    C_D := \frac{\mbox{Average stress at walls} }{\mbox{density}\times(\mbox{wall speed})^2 } = \frac{2}{Re}  D.
\end{equation}
For the constant-$Re$ continuation discussed in this chapter, the drag coefficient differs from the bulk dissipation rate by a constant factor of $2/Re$; so, it is sufficient to plot one of either the drag coefficient or bulk dissipation rate. In keeping with the literature, bulk dissipation rates are shown. 

The kinetic energy density and bulk dissipation rate for flat PCF for the laminar solution and upper and lower branch equilibria are shown in \cref{tab:E-D}.
\begin{table}
    \centering
    \begin{tabular}{c | c | c | c |c}
            & Turb. Mean    & Laminar   & EQ1 (lower)   & EQ2 (upper) \\
            \hline
        $E$   & 0.087         & 0.1667     & 0.1363        & 0.0780    \\
        \hline
        $D$   & 2.926         & 1         & 1.429         & 3.044
    \end{tabular}
    \caption[Energy density and dissipation rate for flat PCF]{Kinetic energy density ($E$) and bulk dissipation rate ($D$) for flat PCF. \label{tab:E-D}}
\end{table}

\begin{figure}
	\centering
	\begin{subfigure}{0.47\textwidth}
		\includegraphics[width=\textwidth]{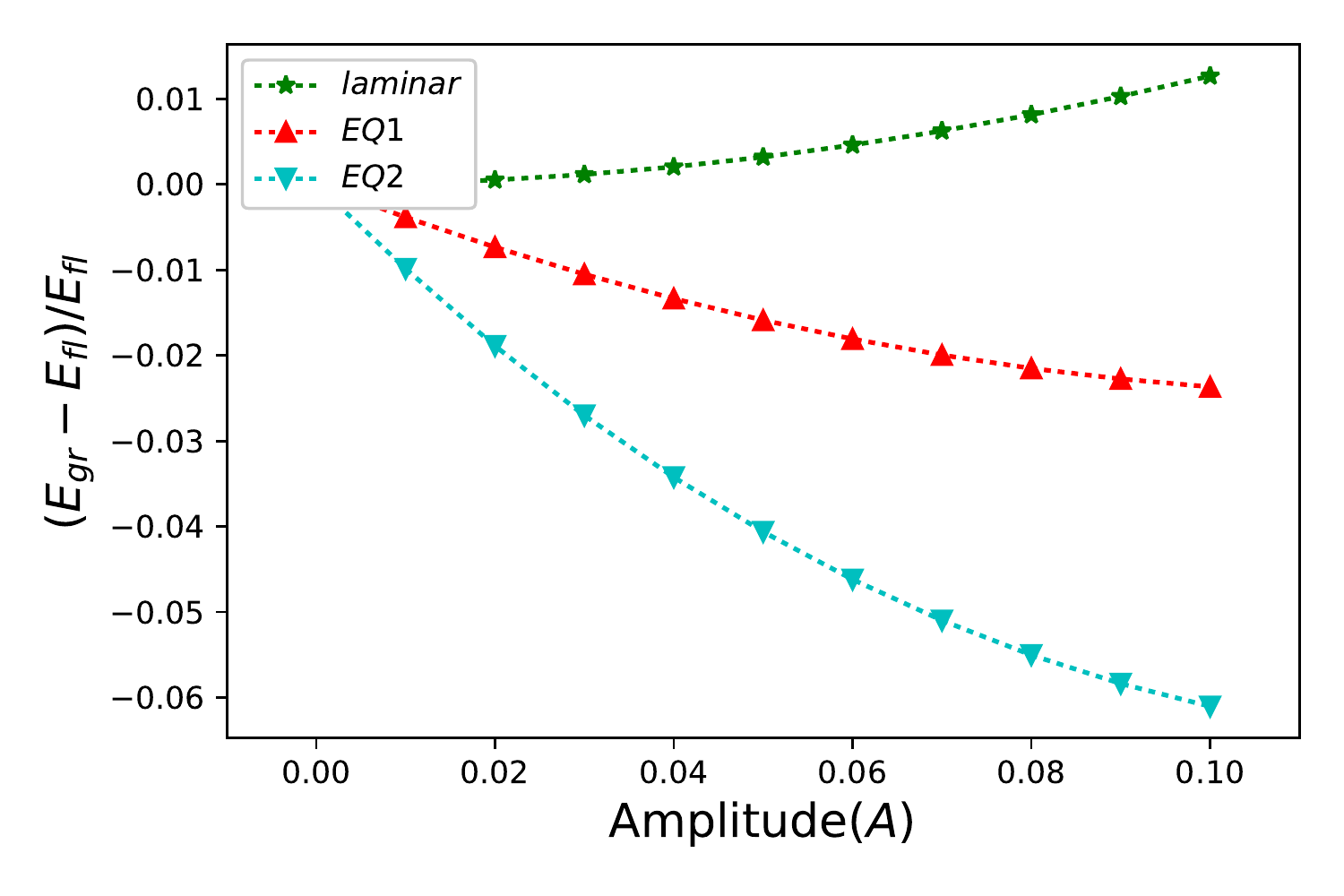}
		\caption{Kinetic energy density; one groove}
	\end{subfigure}
	\begin{subfigure}{0.47\textwidth}
		\includegraphics[width=\textwidth]{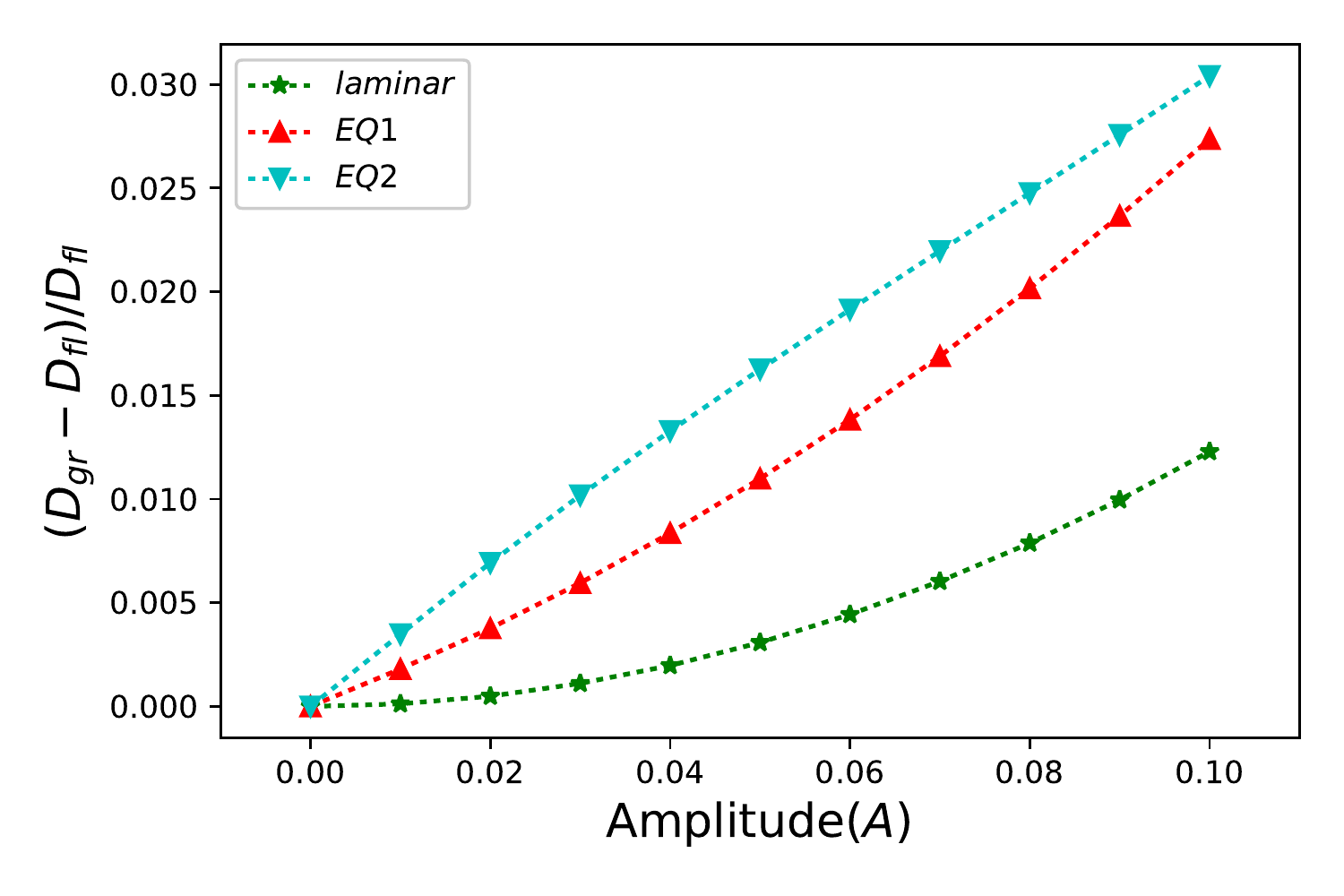}
		\caption{Bulk dissipation rate; one groove}
	\end{subfigure}\\
	\begin{subfigure}{0.47\textwidth}
		\includegraphics[width=\textwidth]{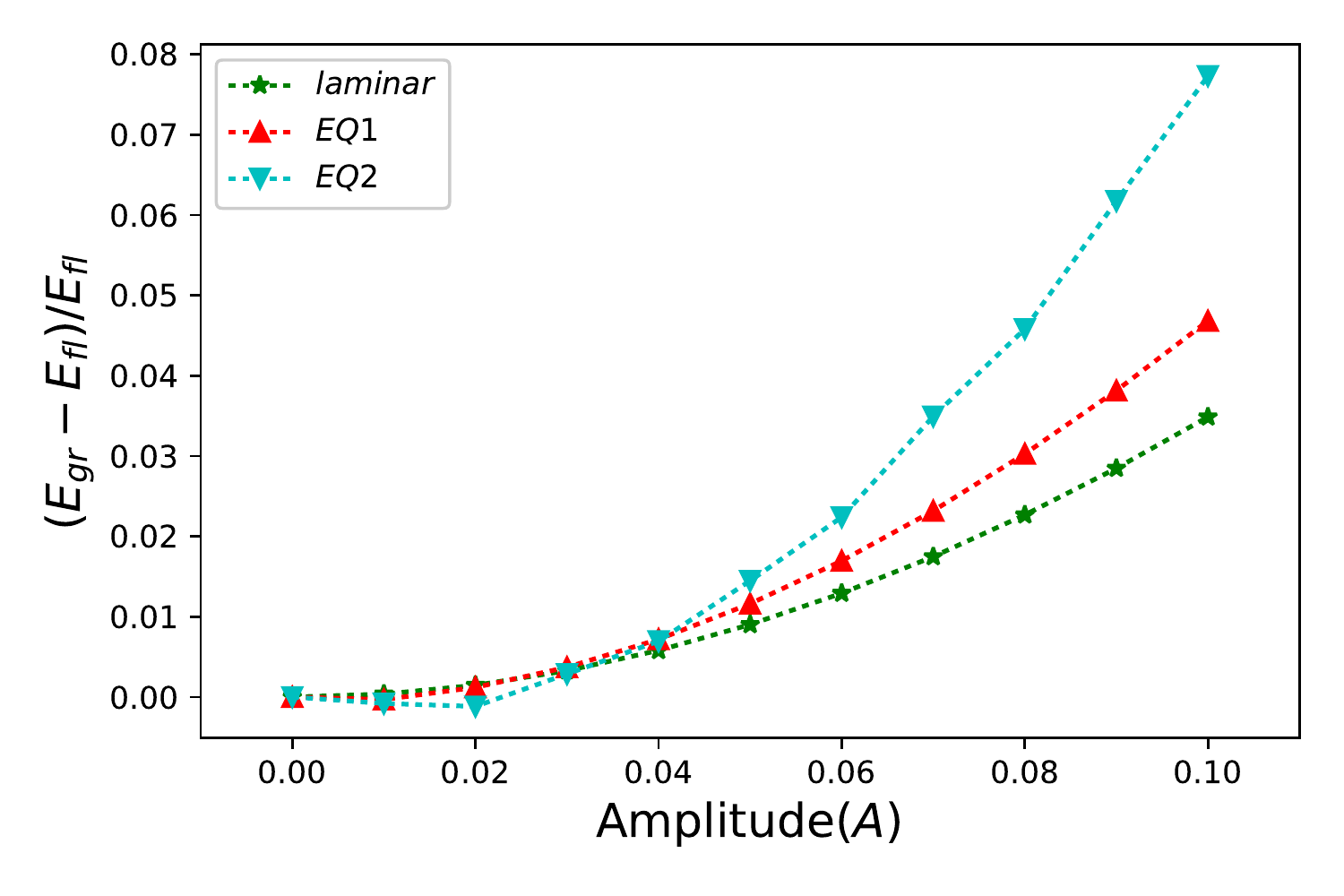}
		\caption{Kinetic energy density; two grooves}
	\end{subfigure}
	\begin{subfigure}{0.47\textwidth}
		\includegraphics[width=\textwidth]{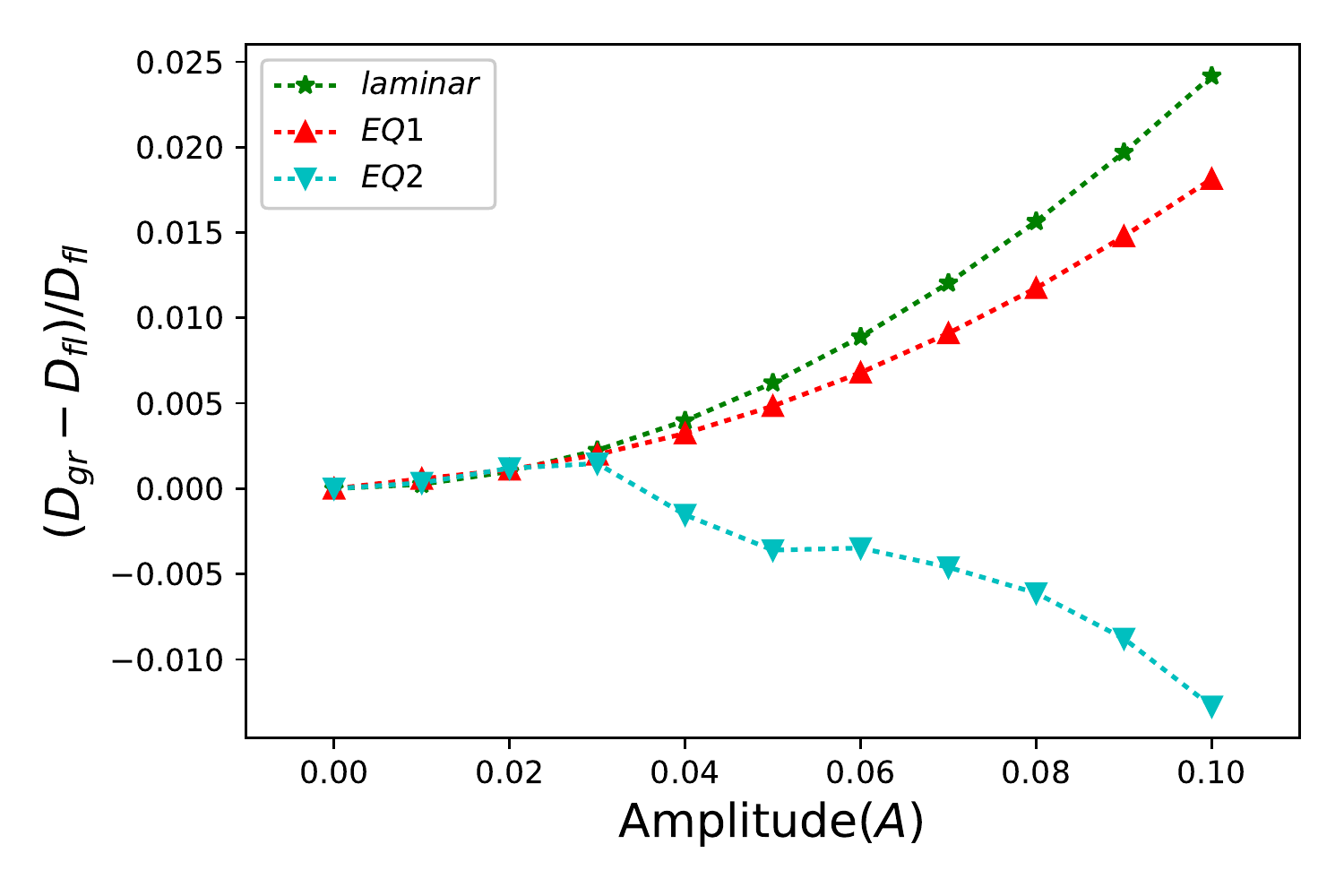}
		\caption{Bulk dissipation rate; two grooves}
	\end{subfigure}\\
	\begin{subfigure}{0.47\textwidth}
		\includegraphics[width=\textwidth]{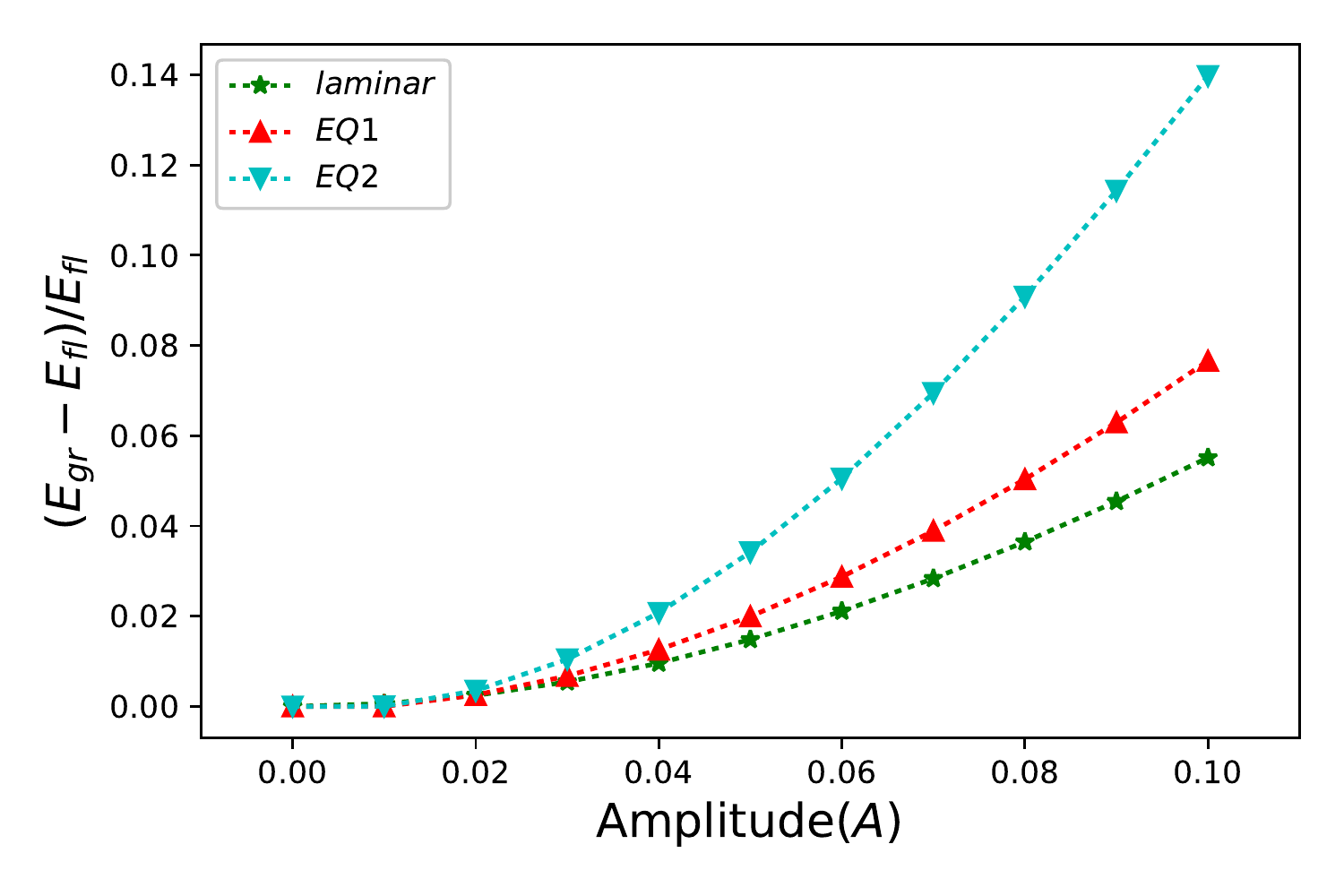}
		\caption{Kinetic energy density; three grooves}
	\end{subfigure}
	\begin{subfigure}{0.47\textwidth}
		\includegraphics[width=\textwidth]{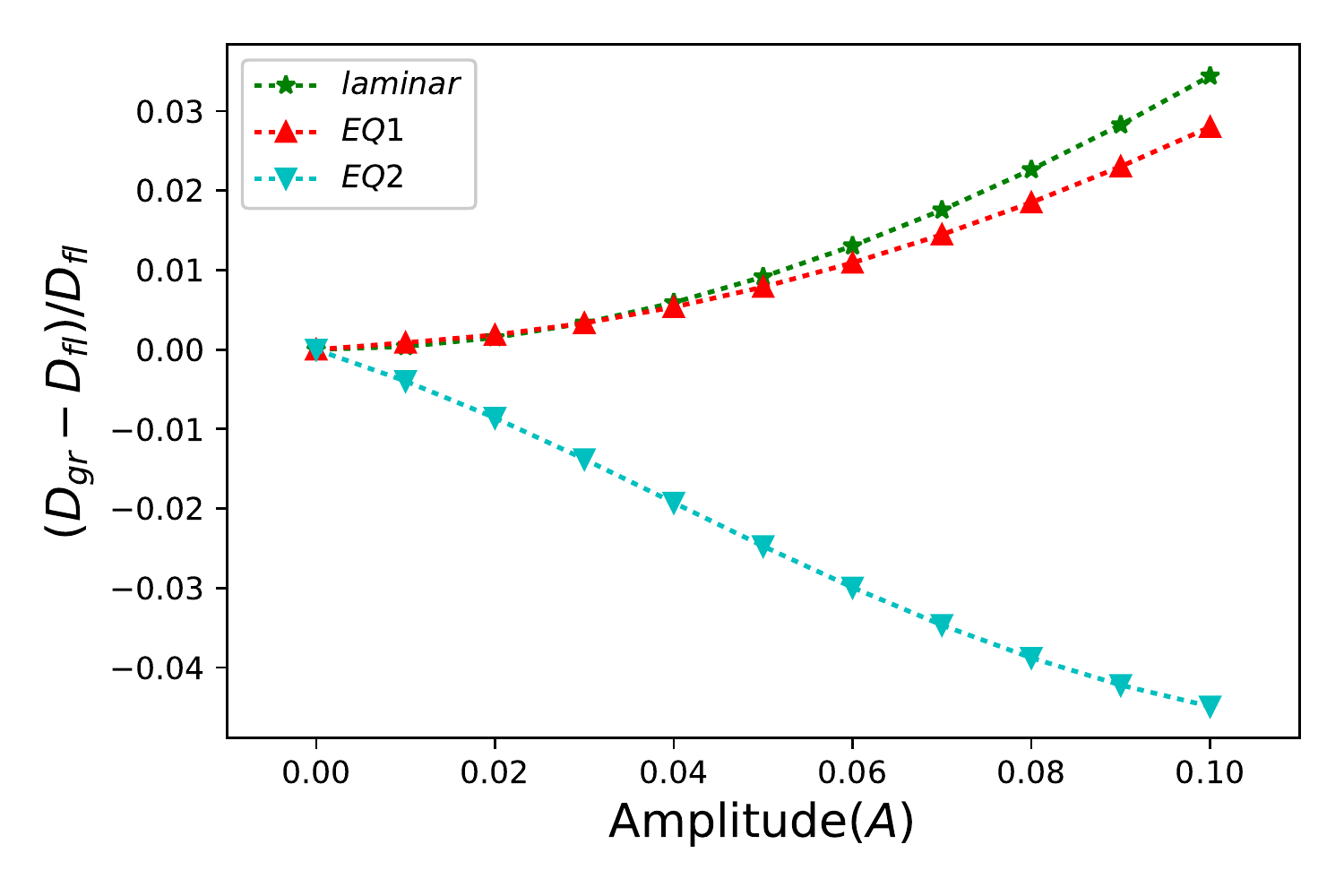}
		\caption{Bulk dissipation rate; three grooves}
	\end{subfigure}\\
	\caption[Kinetic energy density and bulk dissipation rate in grooved PCF with multiple grooves.]{Change in kinetic energy density and bulk dissipation rate in grooved PCF for laminar flow and upper and lower branch equilibria with one, two, and three grooves per box.\label{fig:energy-dissipation-LLU-123} }
\end{figure}

The same quantities for the three wavenumber cases of grooved PCF are shown in \cref{fig:energy-dissipation-LLU-123} for the laminar solution and the equilibria, EQ1 and EQ2. The plots for the one groove per box case are reproduced to facilitate comparison. 

The curves for the laminar solution are discussed first to serve as a base case. All of the curves for the laminar case have zero derivative at $A=0$ because of the highly symmetric nature of the laminar solution. The kinetic energy density and the dissipation rate increase with increasing groove-amplitude, and this increase is stronger for the lower wavelength cases. The increase in kinetic energy is easily understood in terms of the increase in wetted surface area (per unit planform area). The sharper increase in kinetic energy for the three grooves case also follows from the increase in wetted surface area. Bulk dissipation rate, in the laminar case, is mostly due to the wall-normal derivative of the streamwise velocity. Increasing groove amplitude brings high speed regions (of opposite directions) closer, which in turn produces stronger shear and larger dissipation rates.  

We now turn our attention to the kinetic energy density for EQ1 and EQ2 for the three geometries. For the two and three groove cases, the energy density increases, and we can expect this to be for the same reason as the laminar case. Indeed, in \cref{fig:velocity-E1-3,fig:velocity-EQ2-E1-3}, the sizes of the high speed regions (blue at the bottom wall and red at the top wall) do increase as they fill up the area within the grooves. 

The anamolous case is the one groove case, where the energy density decreases with increasing amplitude. This can be explained in terms of the vortex-streak structure ``eating into'' the high speed regions near the wall. The vortex-streak structure is based around the $u=0$ critical layer (it is a critical layer because the equilibria travel with a phase speed of zero). The critical layer moving closer to the wall is accompanied by a shrinkage of the high speed regions, which leads to reduced energy density. It is worth noting that the decrease in energy density for flat PCF from laminar to EQ1 to EQ2 is also linked to the critical layer growing closer to the wall. 

This mechanism for reduction in energy density is most clearly observed by comparing the flat PCF and one groove cases for EQ2 in \cref{fig:velocity-EQ2-E1-3}. At $z\approx 0$ at the bottom wall and $z\approx 1.25$ at the top wall, the critical layer for the one groove case is much closer to the wall than it is for flat PCF. Consequently, there is a significantly smaller high speed region at either locations in the one groove case. In the two and three grooves cases, although the critical layer comes very close to the tips of the grooves, multiple high speed regions exist within the grooves. 

We have observed two phenomena that affect kinetic energy density: (i) increasing wetted surface area leading to increased energy density, and (ii) increased proximity of the critical layer to the walls leading to reduced energy density. Laminar flows in grooved PCF are dominated by the former effect since the critical layer is far from the walls. The one groove case seems to be dominated by the latter effect as the vortex-streak structure penetrates into the grooves. In the two and three grooves cases, the latter effect is not very pronounced as the structure can only reach up to the tips of the grooves, and the increased wetted surface area leads to an overall increase in energy density.

We also note the increase in kinetic energy density for laminar flows to be significantly smaller than that in EQ1 and EQ2 (for two and three grooves per box). This is because the increase in kinetic energy density is normalized by the kinetic energy density of flat PCF, and the kinetic energy density is largest for the laminar case. 

The trends in the bulk dissipation rate can also be explained in terms of the two competing phenomena. Their actions are as follows: 1) Increasing wetted surface area leads to an increase in dissipation rate, and 2) Increasing the proximity of the critical layer to the wall increases the dissipation rate. In laminar flow, as with energy density, the trend in dissipation rate is also dominated by the wetted surface area. 

For EQ2, the critical layer is very close the walls. We expect the trends in dissipation rate to be dominated by the location of the critical layer. For the one groove per box case, increasing amplitude brings the critical layer closer to the walls, leading to increased dissipation rate (as well as power input and drag coefficient). For the three grooves case, increasing amplitude pushes the critical layer outside the tip of the grooves, leading to reduced dissipation rate. For the two grooves case, with increasing amplitude, the dissipation rate first increases and then decreases. However, the dissipation rate for the two grooves case remains lower than that for the one groove case at all amplitudes. This suggests that for the two grooves case, at low amplitudes, the critical layer might penetrate into the grooves. 

For EQ1, the critical layer is far from the walls. The one groove case sees increasing dissipation rate with increasing amplitude as expected. The two and three grooves cases also see an increase in dissipation rate with increasing amplitude, suggesting that the squeezing out of the critical layer from the grooves is either insignificant or absent.  

This mechanism for ``drag reduction'' is consistent with the one discussed in the literature (for instance, see fig. 7 in \citet{lee2001flow}). A Reynolds number dependent optimal spacing of the riblets exists for drag reduction; however, drag reduction is seen for riblet spacing smaller than 25 wall units (\citet{dean2010shark}). The near-wall cycle of turbulence also has a spanwise scale of $\sim 100$ wall-units (for example, \citet{jimenez1991minimal,hamilton1995regeneration}). This suggests that the greatest drag reduction involves riblets whose wavelength is a quarter of the characteristic size of the vortex streak structure. 

The present results show a drag decreasing tendency (for EQ2, which is representative of the near-wall cycle of turbulence) for grooves whose wavelength is a half and a third of the characteristic size of the structure. And indeed, the drag reduction is greater when the riblet spacing is a third of the characteristic size than when it is a half. The present solver, outlined in appendix \ref{app:solver}, cannot solve for cases with smaller wavelengths owing inadequacy in spanwise resolution (limited to 32 grid points), but we can reason towards the observed optimal spacing by considering the two competing phenomena described earlier. Decreasing the wavelength (at a fixed amplitude) produces greater drag due to the increase in wetted surface area. So, the decrease in wavelength is only beneficial until the reduction in drag due to the squeezing out of the critical layer offsets the increase in drag due to the increase in wetted surface area. \Cref{fig:velocity-EQ2-E1-3} shows that the critical layer is already outside the grooves for the three grooves case (i.e., when the wavelength is a third of the characteristic size). Presumably, when the wavelength is reduced beyond a fourth or a fifth of the characteristic size, further decrease in wavelength would not have any significant effect on the critical layer. Beyond this point, reducing the wavelength would only see an increase in drag due to the increase wetted surface area. 

We end this discussion by looking at the groove spacing (wavelength) in terms of inner units, $\lambda_{grooves}Re_\tau$, where $\lambda_{grooves}$ is the wavelength of the grooves (non-dimensionalized by channel half-height) and $Re_\tau$ is the friction Reynolds number. For the one groove per box case, $\lambda_{grooves} = L_z = 0.4$, and for the two and three grooves cases, the groove spacing is $0.2$ and $0.13$ respectively. The friction Reynolds number may be expressed explicitly in terms of the bulk dissipation rate as follows. 
\begin{equation}
        Re_\tau = Re\sqrt{\frac{\hat{u}^2_\tau}{\hat{U}_w^2}}  
                = Re \sqrt{ \frac{1}{2Re} \bigg( \overline{\frac{du}{dy}}\bigg|_{y=-1} + \overline{\frac{du}{dy}}\bigg|_{y=1} \bigg) }  
                = \sqrt{IRe} 
                = \sqrt{DRe}
\end{equation}
where $\hat{u}_\tau$ is the dimensinoal friction velocity, $\hat{U}_w$ is the speed of either walls, $\overline{\frac{du}{dy}}\big|_{y=-1}$ is the average (over plan area) of non-dimensional strain-rate at the bottom wall, $\overline{\frac{du}{dy}}\big|_{y=1}$ is the average (over plan area) of non-dimensional strain-rate at the top wall, and $I$ and $D$ are the power input and bulk dissipation rate respectively. 

For EQ2, the dissipation rate for the different cases lie roughly between $2.9$ and $3.1$, with the flat wall case having $D=3.04$. Using a nominal dissipation rate of $3$, and $Re=400$, the friction Reynolds number is $Re_\tau \approx 35$. Hence, in terms of wall-units, the spacing of the grooves is $14$ for the one groove case, $7$ for the two groove case, and $4.7$ for the three groove case. These spacings seem to be inconsistent with the observations in the literature, especially the drag increase at $14$ inner units. This apparent disparity is easily reconciled by taking into account the fact that the characteristic spanwise size of the vortex-streak structure itself is $14$ inner units, while the characteristic spanwise size of the near-wall cycle in turbulence is about $100$ inner units.

\section{Conclusion}
Equilibria (EQ1 and EQ2) have been successfully continued for PCF from the flat wall case to grooved walls. Discrete symmetries of these equilibria, $s_1$, $s_2$, and $s_3$, have been exploited to allow computations on a 32x35x32 grid (streamwise, wall-normal, spanwise) for geometries with one and three grooves per spanwise period of $2.51$ channel half-heights. The geometry with two grooves per spanwise period (or any even number of grooves per period) does not admit $s_2$ and $s_3$, and this case has been run at a lower resolution of 20x35x32 grid points. Geometries with more than three grooves per periodic box have not been investigated since the resolution with each groove becomes worse with increasing number of grooves. 

The spanwise inhomogeneity of grooved PCF makes the relative spanwise positioning of solutions important. Computations initiated with iterates of non-zero phase fail to convergence, suggesting that such solutions may not even exist. This led us to speculate that symmetric solutions continued to symmetric geometries tend to get spatially anchored about locations that preserve the discrete symmetries. The fate of symmetric solutions when continued to asymmetric geometries remains an open question. 

Grooved PCF with groove-wavelength of a half or a third of the vortex-streak structure being captured have been found to exhibit a drag reducing tendency for upper branch equilibria (EQ2 in the present work). This drag reduction is a result of the squeezing out of the critical layer of the vortex-streak structure by the grooves. If the wavelength is several times lower than this, we expect the drag to increase as the above effect saturates and the wetted surface area becomes the dominating factor. Both phenomena serve to increase the kinetic energy density.

Using the spacing of streamwise streaks in the near-wall cycle ($\approx 100$ wall-units) as the relevant characteristic size, the present results are consistent with the observations of an optimal streak spacing $\approx 25$ wall units. This provides a strong motivation to extend exact invariant solutions to geometries with greater number of grooves per periodic box to find an optimal size for the grooves. However, the limitation imposed by the present numerical method on the resolution stops us from undertaking such an investigation. We hope that future studies involving more efficient numerical scheme for continuation of exact solutions can shed more light on the Reynolds number dependence of optimal riblet spacing.


\textit{Acknowledgement.}---
The authors acknowledge the use of the IRIDIS High Performance Computing Facility, and associated support services at the University of Southampton, in the completion of this work.

\appendix
\section{Steady state solver}\label{app:solver}
The steady-state NSE in the physical domain of $(x,y,z)$ coordinates are given in \cref{governing-equations}. The physical domain is mapped to a flat PCF domain using the domain transformation of \cref{domain-transformation}:
\begin{equation*}
		X = x, \tab
		Y = T(y,z) = y - \sum\limits_m A_m \cos(m\beta z),\tab
		Z  = z.
\end{equation*}
The $z$-derivatives of $T(y,z)$ are expressed as Fourier series in $Z$,
\begin{equation}\label{T-derivatives}
	\begin{aligned}
        T_z(Z) = \sum\limits_{q} T_{z,q} e^{qi\beta Z}   
            &= \sum\limits_{q}  -i q \beta \frac{A_q}{2} e^{iq\beta Z}  \\
		\implies \Aboxed{T_{z,q} &= -i\beta q \frac{A_q}{2} }\\
        T_{zz}(Z)= \sum\limits_{q} T_{zz,q} e^{qi\beta Z}   
			&= \sum\limits_{q}  -(i q \beta)^2 \frac{A_q}{2} e^{iq\beta Z}  \\
		\implies \Aboxed{T_{zz,q} &= \beta^2 q^2 \frac{A_q}{2} }\\
        \bigg(T_z\bigg)^2(Z) = \sum\limits_{q} T^2_{z,q} e^{qi\beta Z} 
            &= \sum\limits_{m} -i m \beta \frac{A_m}{2} e^{im\beta Z}\bigg\{ \sum\limits_q  -i (q-m) \beta \frac{A_{q-m}}{2} e^{i(q-m)\beta z} \bigg\}  \\
        \implies \Aboxed{T^2_{z,q} &= -\beta^2 \sum\limits_m  m(q-m) \frac{A_m A_{q-m}}{4}  }
	\end{aligned}
\end{equation}

The partial derivatives in $(x,y,z)$ relate to partial derivatives in $(X,Y,Z)$ as
\begin{equation}\label{trans-derivatives}
    \begin{gathered}
        \partial_x = \partial_X, \tab \partial_{xx} = \partial_{XX},\tab
        \partial_y = \partial_Y, \tab \partial_{yy} = \partial_{YY},\\
        \partial_z = \partial_Z + T_z \partial_Y, \tab
        \partial_{zz} = \partial_{ZZ} + 2T_z \partial_{YZ} + T_{zz} \partial_Y + (T_z)^2 \partial_{YY}.
    \end{gathered}
\end{equation}

The variables, $u$, $v$, $w$, and $p$ are discretized in the transformed domain (Fourier spectral method in $x$ and $z$, and Chebyshev collocation method in $y$). The spatial derivatives in the physical domain in the steady-state NSE are expressed in terms of derivatives in the transformed domain; the Fourier coefficients for each term in the governing equation are now presented. Fourier coefficient of $u$ for the mode $e^{i(l\alpha X + m\beta Z)}$ is denoted $\{u\}_{l,m}$ or $u_{l,m}$, and the $Y$-dependence is not explicitly shown. 

Fourier coefficients for the pressure gradient terms are
\begin{equation*}
\begin{aligned}
    \big\{\partial_{x} p\big\}_{l,m} &= \big\{ \partial_{X}  p\big\}_{l,m}                  =  il\alpha p_{l,m}, \\
    \big\{\partial_{y} p\big\}_{l,m} &= \big\{ \partial_{Y}  p\big\}_{l,m}                  =  Dp_{l,m}, \\
    \big\{\partial_{z} p\big\}_{l,m} &= \big\{ \big(\partial_{Z}+T_z \partial_Y\big) p\}_{l,m}   =  im\beta p_{l,m} + \sum\limits_q T_{z,q} Dp_{l,m-q}. \\
\end{aligned}
\end{equation*}
where $D$ is a linear operator representing differentiation with respect to $Y$, and $T_{z,q}$ is defined in \cref{T-derivatives}.

The Fourier coefficients of the diffusion term, involving the Laplacian of the velocity, can similarly be expressed in terms of Fourier coefficients of the velocity field. The diffusion term is $\frac{1}{Re} \Delta u^i$; the Fourier coefficients of the Laplacian of streamwise velocity are 
\begin{equation} \label{diff}
    \begin{aligned}
        \big\{\partial_{x_jx_j}u \big\}_{l,m} 
                =&  \big(-l^2 \alpha^2 - m^2 \beta^2 + D^2\big) u_{l,m}\\
                    &+ \sum\limits_q \big(2i(m-q)\beta T_{z,q} D + T_{zz,q}D + T^2_{z,q}D^2     \big) u_{l,m-q},
    \end{aligned}
\end{equation}
using Einstein summation notation for $\partial_{x_jx_j}$, $j \in \{1,2,3\}$, with $(x_1,x_2,x_3)=(x,y,z)$. The Laplacians of the wall-normal and spanwise velocity involve similar expressions for their Fourier coefficients. 

The convection term is fully non-linear; the velocity is not decomposed into a base-flow and fluctuations. The Fourier coefficients of the convection term in the streamwise momentum equation are expressed using sums of products of Fourier coefficients of the velocity field as
\begin{equation}\label{conv}
\begin{gathered}
    \bigg\{ \big(u \partial_{x}+v\partial_y + w\partial_z\big) u \bigg\}_{l,m}		
        = \sum\limits_{l'} \sum\limits_{m'} \bigg\{
             i(l-l')\alpha u_{l',m'} u_{l-l',m-m'}
            +           v_{l',m'} D  u_{l-l',m-m'}\\
            +i(m-m')\beta  w_{l',m'} u_{l-l',m-m'}
            + \sum\limits_q  T_{z,q} w_{l',m'} D u_{l-l',m-m'-q}
        \bigg\}.
\end{gathered}
\end{equation}
Note the extra terms in the triadic interactions which are restricted to modes of the form $\big\{(l,m), (l',m'),(l-l',m-m')\big\}$ in flat PCF. The terms in the wall-normal and spanwise momentum equations have similar expressions.

The continuity equation is also projected onto the Fourier modes as
\begin{equation} \label{cont}
    0=\big\{ \nabla \cdot \boldsymbol{u} \big\}_{l,m} =  
       il\alpha u_{l,m} + Dv_{l,m} + im\beta w_{l,m} + \sum\limits_q  T_{z,q} Dw_{l,m-q}
\end{equation}

In the transformed system, the walls are at $Y=\pm 1$. Dirichlet conditions are imposed on each Fourier coefficient for the velocities and the divergence as 
\begin{equation}\label{eq:BCs}
    \begin{aligned}
        u_{l,m}(Y=\pm 1) &= 0, \quad v_{l,m}(Y=\pm 1) &= 0, \\ 
        w_{l,m}(Y=\pm1)  &=0, \quad
        \big\{\nabla \cdot \boldsymbol{u} \big\}_{l,m} (Y=\pm 1) &= 0 .
    \end{aligned}
\end{equation}
However, the boundary conditions on $u_{l=0,m=0}$ are set as $u_{l=0,m=0}(Y=\pm1) = \pm 1$ to allow for wall motion.

\subsection{Matrix equation and Iterative scheme}
The discretized steady-state NSE are cast as a non-linear matrix equation 
\begin{equation}\label{governing-discrete}
	\boldsymbol{F}(\boldsymbol{\chi}) := \mathcal{L}\boldsymbol{\chi} + \boldsymbol{\mathcal{N}}(\boldsymbol{\chi}) - \boldsymbol{f} = 0, 
\end{equation}
where $\boldsymbol{\chi}$ is the state-vector containing the Fourier coefficients of $u$, $v$, $w$, and $p$, $\mathcal{L}$ is a matrix representing the action of the linear terms, $\boldsymbol{\mathcal{N}}(\boldsymbol{\chi})$ represents the non-linear term (convection), and $\boldsymbol{f}$ represents a forcing. The forcing term is used to impose boundary conditions (or mean pressure gradient if a channel flow is to be solved for).

The state-vector $\boldsymbol{\chi}$ is built as 
\begin{equation}\label{state-order}
	\begin{gathered}
		\boldsymbol{\chi} = \begin{bmatrix} \chi_{-L,-M}\\ \chi_{-L,-M+1} \\ \vdots \\ \chi_{-L,M}\\ \chi_{-L,-M}\\ \vdots \\ \chi_{L,M} \end{bmatrix}, 
            \tab \mbox{where,    } \chi_{l,m} = \begin{bmatrix} u_{l,m}(Y) \\ v_{l,m}(Y) \\ w_{l,m}(Y) \\ p_{l,m}(Y) \end{bmatrix},
	\end{gathered}
\end{equation}
where $L$ and $M$ are positive integers representing the streamwise and spanwise resolutions, and $\chi_{l,m}$ is used to represent the coefficients for a particular Fourier mode. The entries of $u_{l,m}(Y)$ and other components extend from $Y=-1$ to $Y=1$, including the boundaries. Each Fourier coefficient ($\chi_{l,m}$) has $4N$ entries, and $4N$ equations are solved. $4(N-2)$ of these equations are the NSE (including continuity equation) at the internal nodes; each row of $\mathcal{L} + \boldsymbol{\mathcal{N}}-\boldsymbol{f}$ corresponds to one out of the four NSE imposed at one internal wall-normal location for one Fourier mode. The remaining 8 rows are used for the boundary conditions of \cref{eq:BCs}. To impose boundary conditions, all entries of $\boldsymbol{\mathcal{N}}$ corresponding to the 8 boundary rows are set to zero, and the entries of $\mathcal{L}$ are set to reflect the terms on the left in each of the boundary conditions in \cref{eq:BCs}. The terms on the right in the boundary conditions in \cref{eq:BCs}, either 0 or 1, are set in the forcing term, $\boldsymbol{f}$.  

\subsubsection{Newton-Raphson method}
The matrix equation is solved using the exact Newton-Raphson method, which iteratively refines a given estimate $\boldsymbol{\chi}^0$ of an exact solution $\boldsymbol{\chi}^*$ as
\begin{equation}\label{iter-scheme}
\begin{aligned}
	\boldsymbol{\chi}^{m+1} &= \boldsymbol{\chi}^m - \big\{\boldsymbol{F}'(\boldsymbol{\chi}^m)\big\}^{-1} \boldsymbol{F}(\boldsymbol{\chi}^m)\\
		&= \boldsymbol{\chi}^m - \big\{\mathcal{L} + \mathcal{G}(\boldsymbol{\chi}^m)\big\}^{-1} \boldsymbol{F}(\boldsymbol{\chi}^m).
\end{aligned}
\end{equation}
until a convergence criterion is satisfied. Here, $\boldsymbol{F}'$ denotes the Jacobian of $\boldsymbol{F}$, and the state-dependent matrix $\mathcal{G}(\boldsymbol{\chi})$ is the Jacobian of only the non-linear term $\boldsymbol{\mathcal{N}}(\boldsymbol{\chi})$; $\mathcal{G}(\boldsymbol{\chi})$ can be shown to relate to the non-linear term as 
\begin{equation}\label{jacobian-convection}
	\boldsymbol{\mathcal{N}}(\boldsymbol{\chi}) = \frac{1}{2} \mathcal{G}(\boldsymbol{\chi}) \boldsymbol{\chi}.
\end{equation}
The superscript $m$ on $\boldsymbol{\chi}$ is the iteration number, while subscripts in \cref{state-order} represent Fourier coefficients of the fields at a particular iteration. 

Preliminary numerical experiments with computing laminar solutions for grooved PCF and channel flows showed no improvement in convergence rate using parametric continuation. Hence, the flat PCF solution, either EQ1 or EQ2, is used as the initial estimate for all grooved PCF cases; the transformed domain of $(X,Y,Z)$ coordinates where the variables are discretized is identical to the flat PCF domain. 

The Jacobian is formed (i.e. stored in memory) and inverted using an SVD-based inversion routine (lstsq of SciPy, which wraps around LAPACK's gelsd). This iterative scheme is much simpler, and specific, compared to that of \cite{channelflow} and \cite{viswanath2007recurrent}, whose state equation is based on a time-$t$ map instead of the steady-state NSE; in their case, the Jacobian of $\boldsymbol{F}$ is only expressed in terms of its action as a matrix-vector product, and a matrix-free variant of the Newton-Raphson method involving a constrained hookstep is used. The resolutions used in the present work are limited due to the explicit storage and inversion of the Jacobian; nonetheless, it allows simple computations that can be verified by computing the residual norm of the solutions on finer grids, which is computationally cheap. 

To improve convergence of the Newton iterations, a line search method is used at the end of each iteration. If $\delta \boldsymbol{\chi}^+ = -\boldsymbol{F}'^{-1}(\chi) \boldsymbol{F}(\boldsymbol{\chi})$ is the solution to the matrix equation, then an optimal correction, $\delta \boldsymbol{\chi}^*$ along $\delta\boldsymbol{\chi}^+$, is found using a binary search over $d_f$ as
\begin{equation}
	\delta \boldsymbol{\chi}^{*}= 
	\{\underset{d_f \in \mathbb{R}}{\textrm{arg\,min}} ||\boldsymbol{F}(\boldsymbol{\chi}^+ + d_f \delta \boldsymbol{\chi}^{+})||\} \, \delta\boldsymbol{\chi}^+,
	\mbox{    with } \delta \boldsymbol{\chi}^+ =- \{\boldsymbol{F}'\}^{-1}\boldsymbol{F}(\boldsymbol{\chi}). 
\end{equation}

The convergence criterion is that the residual norm,
\begin{equation}\label{residual-norm}
    r =  \sqrt{\sum\limits_{l=-L}^L \sum\limits_{m=-M}^M \frac{1}{2} \int_{Y=-1}^{1} \big\{ |\partial_t u_{l,m}|^2+ |\partial_t v_{l,m}|^2+|\partial_t w_{l,m}|^2+|(\nabla \cdot \boldsymbol{u})|_{l,m}^2 \big\}\quad dY}
\end{equation}
goes below a prescribed tolerance. The tolerance was usually set to $10^{-12}$, but this could not be reached in some cases (two groove per box). To ensure that the steady solutions are grid-independent, an spatial accuracy ($r_{2x}$) is defined as the residual obtained when the state-vector is interpolated to a grid of twice the number of nodes(/modes) in each coordinate direction. 

The divergence-free condition is used directly as an equation in $\boldsymbol{F}(\boldsymbol{\chi})=0$, and the iterations converge without issue. Since the algorithm does not involve time-marching, we found no particular reason to use a pressure Poisson equation instead of the divergence equation for velocity.

\bibliographystyle{jfm}

\bibliography{manuscript}

\begin{thebibliography}{30}
\expandafter\ifx\csname natexlab\endcsname\relax\def\natexlab#1{#1}\fi
\def\au#1{#1} \def\ed#1{#1} \def\yr#1{#1}\def\at#1{#1}\def\jt#1{\textit{#1}}
  \def\bt#1{#1}\def\bvol#1{\textbf{#1}} \def\vol#1{#1} \def\pg#1{#1}
  \def\publ#1{#1}\def\arxiv#1{#1}\def\org#1{#1}\def\st#1{\textit{#1}}

\bibitem[Bannier {\em et~al.\/}(2016)Bannier, Garnier \&
  Sagaut]{bannier2016riblets}
{\sc \au{Bannier, A.}, \au{Garnier, E.} \& \au{Sagaut, P.}} \yr{2016}
  \at{Riblets induced drag reduction on a spatially developing turbulent
  boundary layer}.  \bt{In {\em Progress in Wall Turbulence 2\/}},  \pg{pp.
  213--224}.  \publ{Springer}.

\bibitem[Blackburn {\em et~al.\/}(2013)Blackburn, Hall \&
  Sherwin]{blackburn2013lower}
{\sc \au{Blackburn, H.~M.}, \au{Hall, P.} \& \au{Sherwin, S.~J.}} \yr{2013}
  \at{Lower branch equilibria in couette flow: the emergence of canonical
  states for arbitrary shear flows}.  \jt{Journal of Fluid Mechanics}
  \bvol{726},  \pg{R2}.

\bibitem[Chantry \& Kerswell(2015)]{chantry2015localization}
{\sc \au{Chantry, M.} \& \au{Kerswell, R.~R.}} \yr{2015}  \at{Localization in a
  spanwise-extended model of plane couette flow}.  \jt{Physical Review E}
  \bvol{91}~(4),  \pg{043005}.

\bibitem[Dean \& Bhushan(2010)]{dean2010shark}
{\sc \au{Dean, B.} \& \au{Bhushan, B.}} \yr{2010}  \at{Shark-skin surfaces for
  fluid-drag reduction in turbulent flow: a review}.  \jt{Philosophical
  Transactions of the Royal Society of London A: Mathematical, Physical and
  Engineering Sciences}  \bvol{368}~(1929),  \pg{4775--4806}.

\bibitem[Eckhardt {\em et~al.\/}(2008)Eckhardt, Faisst, Schmiegel \&
  Schneider]{eckhardt2008dynamical}
{\sc \au{Eckhardt, B.}, \au{Faisst, H.}, \au{Schmiegel, A.} \& \au{Schneider,
  T.~M.}} \yr{2008}  \at{Dynamical systems and the transition to turbulence in
  linearly stable shear flows}.  \jt{Philosophical Transactions of the Royal
  Society of London A: Mathematical, Physical and Engineering Sciences}
  \bvol{366}~(1868),  \pg{1297--1315}.

\bibitem[Flack \& Schultz(2014)]{flack2014roughness}
{\sc \au{Flack, K.~A.} \& \au{Schultz, M.~P.}} \yr{2014}  \at{Roughness effects
  on wall-bounded turbulent flows a)}.  \jt{Physics of Fluids (1994-present)}
  \bvol{26}~(10),  \pg{101305}.

\bibitem[Gibson \& Cvitanovi{\'c}(2008)]{gibson2008visualizing}
{\sc \au{Gibson, J. F.and~Halcrow, J.} \& \au{Cvitanovi{\'c}, P.}} \yr{2008}
  \at{Visualizing the geometry of state space in plane couette flow}.
  \jt{Journal of Fluid Mechanics}  \bvol{611},  \pg{107--130}.

\bibitem[Gibson(2014)]{channelflow}
{\sc \au{Gibson, J.~F.}} \yr{2014}  \bt{{Channelflow}: {A} spectral
  {Navier-Stokes} simulator in {C}++}. {\em Tech. Rep.\/}.  \org{U. New
  Hampshire}, {\tt {Channelflow.org}}.

\bibitem[Gibson \& Brand(2014)]{gibson2014spanwise}
{\sc \au{Gibson, J.~F.} \& \au{Brand, E.}} \yr{2014}  \at{Spanwise-localized
  solutions of planar shear flows}.  \jt{Journal of Fluid Mechanics}
  \bvol{745},  \pg{25--61}.

\bibitem[Gibson {\em et~al.\/}(2009)Gibson, Halcrow \&
  Cvitanovi{\'c}]{gibson2009equilibrium}
{\sc \au{Gibson, John~F}, \au{Halcrow, Jonathan} \& \au{Cvitanovi{\'c},
  Predrag}} \yr{2009}  \at{Equilibrium and travelling-wave solutions of plane
  couette flow}.  \jt{Journal of Fluid Mechanics}  \bvol{638},  \pg{243--266}.

\bibitem[Halcrow(2008)]{halcrow2008charting}
{\sc \au{Halcrow, Jonathan~J}} \yr{2008}  \at{Charting the state space of plane
  couette flow: equilibria, relative equilibria, and heteroclinic connections}.
  PhD thesis, Georgia Institute of Technology.

\bibitem[Hall \& Sherwin(2010)]{hall2010streamwise}
{\sc \au{Hall, P.} \& \au{Sherwin, S.}} \yr{2010}  \at{Streamwise vortices in
  shear flows: harbingers of transition and the skeleton of coherent
  structures}.  \jt{Journal of Fluid Mechanics}  \bvol{661},  \pg{178--205}.

\bibitem[Hall \& Smith(1991)]{hall1991strongly}
{\sc \au{Hall, P.} \& \au{Smith, F.~T.}} \yr{1991}  \at{On strongly nonlinear
  vortex/wave interactions in boundary-layer transition}.  \jt{Journal of Fluid
  Mechanics}  \bvol{227},  \pg{641--666}.

\bibitem[Hamilton {\em et~al.\/}(1995)Hamilton, Kim \&
  Waleffe]{hamilton1995regeneration}
{\sc \au{Hamilton, J.~M.}, \au{Kim, J.} \& \au{Waleffe, F.}} \yr{1995}
  \at{Regeneration mechanisms of near-wall turbulence structures}.  \jt{Journal
  of Fluid Mechanics}  \bvol{287},  \pg{317--348}.

\bibitem[Jim{\'e}nez(2004)]{jimenez2004turbulent}
{\sc \au{Jim{\'e}nez, J.}} \yr{2004}  \at{Turbulent flows over rough walls}.
  \jt{Annual Review of Fluid Mechanics}  \bvol{36},  \pg{173--196}.

\bibitem[Jim{\'e}nez \& Moin(1991)]{jimenez1991minimal}
{\sc \au{Jim{\'e}nez, J.} \& \au{Moin, P.}} \yr{1991}  \at{The minimal flow
  unit in near-wall turbulence}.  \jt{Journal of Fluid Mechanics}  \bvol{225},
  \pg{213--240}.

\bibitem[Kasliwal {\em et~al.\/}(2012)Kasliwal, Duncan \&
  Papachristodoulou]{kasliwal2012modelling}
{\sc \au{Kasliwal, A.}, \au{Duncan, S.} \& \au{Papachristodoulou, A.}}
  \yr{2012} Modelling channel flow over riblets: Calculating the energy
  amplification.  \bt{In {\em Control (CONTROL), 2012 UKACC International
  Conference on\/}},  \pg{pp. 625--630}. IEEE.

\bibitem[Kawahara {\em et~al.\/}(2012)Kawahara, Uhlmann \& van
  Veen]{kawahara2012significance}
{\sc \au{Kawahara, G.}, \au{Uhlmann, M.} \& \au{van Veen, L.}} \yr{2012}
  \at{The significance of simple invariant solutions in turbulent flows}.
  \jt{Annual Review of Fluid Mechanics}  \bvol{44},  \pg{203--225}.

\bibitem[Lee \& Lee(2001)]{lee2001flow}
{\sc \au{Lee, S-J} \& \au{Lee, S-H}} \yr{2001}  \at{Flow field analysis of a
  turbulent boundary layer over a riblet surface}.  \jt{Experiments in Fluids}
  \bvol{30}~(2),  \pg{153--166}.

\bibitem[Moradi \& Floryan(2013)]{moradi2013maximization}
{\sc \au{Moradi, H.~V.} \& \au{Floryan, J.~M.}} \yr{2013}  \at{Maximization of
  heat transfer across micro-channels}.  \jt{International Journal of Heat and
  Mass Transfer}  \bvol{66},  \pg{517--530}.

\bibitem[Nagata(1990)]{nagata1990three}
{\sc \au{Nagata, M.}} \yr{1990}  \at{Three-dimensional finite-amplitude
  solutions in plane couette flow: bifurcation from infinity}.  \jt{Journal of
  Fluid Mechanics}  \bvol{217},  \pg{519--527}.

\bibitem[Park \& Graham(2015)]{park2015exact}
{\sc \au{Park, J.~S.} \& \au{Graham, M.~D.}} \yr{2015}  \at{Exact coherent
  states and connections to turbulent dynamics in minimal channel flow}.
  \jt{Journal of Fluid Mechanics}  \bvol{782},  \pg{430--454}.

\bibitem[Perry \& Maru{\v{s}}ic(1995)]{perry1995wall}
{\sc \au{Perry, A.~E.} \& \au{Maru{\v{s}}ic, I.}} \yr{1995}  \at{A wall-wake
  model for the turbulence structure of boundary layers. part 1. extension of
  the attached eddy hypothesis}.  \jt{Journal of Fluid Mechanics}  \bvol{298},
  \pg{361--388}.

\bibitem[Squire {\em et~al.\/}(2016)Squire, Morrill-Winter, Hutchins, Schultz,
  Klewicki \& Marusic]{squire2016comparison}
{\sc \au{Squire, D.~T.}, \au{Morrill-Winter, C.}, \au{Hutchins, N.},
  \au{Schultz, M.~P.}, \au{Klewicki, J.~C.} \& \au{Marusic, I.}} \yr{2016}
  \at{Comparison of turbulent boundary layers over smooth and rough surfaces up
  to high reynolds numbers}.  \jt{Journal of Fluid Mechanics}  \bvol{795},
  \pg{210--240}.

\bibitem[Townsend(1980)]{townsend1980structure}
{\sc \au{Townsend, A.~A.}} \yr{1980} {\em The structure of turbulent shear
  flow\/}.  \publ{Cambridge university press}.

\bibitem[Viswanath(2007)]{viswanath2007recurrent}
{\sc \au{Viswanath, D.}} \yr{2007}  \at{Recurrent motions within plane couette
  turbulence}.  \jt{Journal of Fluid Mechanics}  \bvol{580},  \pg{339--358}.

\bibitem[Waleffe(1997)]{waleffe1997self}
{\sc \au{Waleffe, F.}} \yr{1997}  \at{On a self-sustaining process in shear
  flows}.  \jt{Physics of Fluids (1994-present)}  \bvol{9}~(4),  \pg{883--900}.

\bibitem[Walsh(1990)]{walsh1990effect}
{\sc \au{Walsh, M.~J.}} \yr{1990}  \at{Effect of detailed surface geometry on
  riblet drag reduction performance}.  \jt{Journal of Aircraft}  \bvol{27}~(6),
   \pg{572--573}.

\bibitem[Willis {\em et~al.\/}(2013)Willis, Cvitanovi{\'c} \&
  Avila]{willis2013revealing}
{\sc \au{Willis, A.~P.}, \au{Cvitanovi{\'c}, P.} \& \au{Avila, M.}} \yr{2013}
  \at{Revealing the state space of turbulent pipe flow by symmetry reduction}.
  \jt{Journal of Fluid Mechanics}  \bvol{721},  \pg{514--540}.

\bibitem[Willis {\em et~al.\/}(2017)Willis, Short, Burak~Budanur, Farazmand \&
  Cvitanovic]{willis2015relative}
{\sc \au{Willis, A.~P.}, \au{Short, K.~Y.}, \au{Burak~Budanur, N.},
  \au{Farazmand, M.} \& \au{Cvitanovic, P.}} \yr{2017}  \at{Relative periodic
  orbits form the backbone of turbulent pipe flow}.  \jt{arXiv preprint
  arXiv:1705.03720} .

\end{thebibliography}

\end{document}